\documentclass[article]{aa}
\usepackage[varg]{txfonts}
\usepackage{color}
\usepackage{epsfig,graphicx,lscape,subfigure}
\usepackage{float,longtable,comment}
\usepackage{rotating,amssymb,amsmath}
\usepackage{natbib}

\begin{document}

   \title{Luminosity functions of cluster galaxies}

\titlerunning{NUV LFs of Cluster Galaxies}
   \subtitle{The Near-ultraviolet luminosity function at $<z> \sim 0.05$}
  \author{Roberto De Propris\inst{\ref{i1}}, Malcolm N. Bremer\inst{\ref{i2}}, Steven Phillipps\inst{\ref{i2}}}
\institute{FINCA, University of Turku, Vesilinnantie 5, 20014, Turku, Finland\label{i1}
    \and{H. H. Wills Physics Laboratory, University of Bristol, Tyndall Avenue, Bristol, BS1 8TL, UK}\label{i2}
}
   \date{}

 
  \abstract{
We derive NUV luminosity functions for 6471 NUV detected galaxies in 28 $0.02 < z < 0.08$
clusters and consider their dependence on cluster properties. We consider optically 
red and blue galaxies and explore how their NUV LFs vary in several cluster 
subsamples, selected to best show the influence of environment. Our composite 
LF is well fit by the Schechter form with $M^*_{NUV}=-18.98 \pm 0.07$ and $\alpha=-1.87 \pm 0.03$ 
in good agreement with values for the Coma centre and the Shapley supercluster,  but with a 
steeper slope and brighter $L^*$ than in Virgo. The steep slope 
is due to the contribution of massive quiescent galaxies that are faint 
in the NUV. There are significant differences in the NUV LFs for clusters having low and high  
X-ray luminosities and for sparse and dense clusters, though none are particularly well fitted 
by the Schechter form, making a physical interpretation of the parameters difficult. When splitting
clusters into two subsamples by X-ray luminosity, the ratio of low to high NUV luminosity galaxies 
is higher in the high X-ray luminosity subsample  (i.e the luminosity function is steeper across 
the sampled luminosity range). In subsamples split by surface density, when characterised by 
Schechter functions the dense clusters have an $M^*$  about a magnitude fainter than that of the  
sparse clusters and $\alpha$ is steeper ($-1.9$ vs. $-1.6$ respectively). The differences in the 
data appear to be driven by changes in the LF of blue (star-forming) galaxies. This appears to be 
related to interactions with the cluster gas.  For the blue galaxies alone, the luminosity 
distributions indicate that for high $L_X$ and high velocity dispersion cluster subsamples (i.e. 
the higher mass clusters), there are relatively fewer high UV luminosity galaxies (or 
correspondingly a relative excess of low UV luminosity galaxies) in comparison the lower mass 
cluster subsamples.}


   \maketitle
%

\section{Introduction}

The luminosity function (hereafter LF) of a population of galaxies characterises the end result of 
a series of processes involving the mass assembly, star formation and morphological evolution of 
these galaxies. The LF therefore provides a statistical first-order description of the properties of 
a population of galaxies at a given epoch and yields a test of models for galaxy formation and 
evolution and assembly \citep[e.g.][]{henriques2015,rodrigues2017,mitchell2018}. 

LFs measured in different bandpasses measure, and are sensitive to, different mechanisms involved 
in galaxy evolution. As an example, infrared LFs more closely resemble  stellar mass functions and 
can be used to determine the mass assembly history of galaxies. In contast, the near ultraviolet 
(NUV) LF is instead strongly affected by star formation and its evolution (and dependence on 
environment) as most of the flux is produced by young stars with relatively short lifetimes ($< 1$ 
Gyr), although significant NUV flux can also be produced by hot horizontal branch stars in old 
stellar populations. Therefore, NUV observations are expected to provide clues to the importance 
of environmental effects in quenching star formation, especially within the cores of galaxy clusters 
and other high density regions, where a variety of processes (e.g., tides, interactions, mergers, 
ram stripping) are believed to play an important role in controlling star formation 
\citep[e.g][]{boselli2014,taranu2014,boselli2016a,darvish2018}.

It is only recently that panoramic images in the vacuum UV have become available from the GALEX 
telescope \citep{martin2005}. Limited areas of sky were earlier observed from balloon-borne 
telescopes or experiments flown on the Space Shuttle. GALEX has provided us with the first 
nearly all-sky survey in the vacuum UV bands at 1500 and 2300 \AA. From this, \cite{wyder2005} 
produced the first (and so far, only) field UV luminosity function, but only a few clusters have 
been so characterised until now. From FOCA data \cite{andreon1999} derived a 2000 \AA\ LF for the 
Coma cluster, while \cite{cortese2003} obtained a composite LF for Virgo, Coma and Abell 1367 
(also at 2000 \AA). Based on GALEX data (but excluding the centre) \cite*{cortese2008} determined 
the UV LFs for galaxies in the Coma cluster, including quiescent and star-forming galaxies separately. 
Virgo has been extensively studied in the same manner by the GUVICS collaboration \citep{boselli2016b},
while \cite{hammer2012} have obtained a LF for a field in Coma lying between the core and the 
infalling NGC4839 subgroup. Finally, \cite{haines2011} measured a composite UV LF for five clusters 
within the Shapley supercluster of galaxies.

The characteristic magnitude $L^*$ is observed to vary widely between Virgo, Coma, A1367 and 
the composite Shapley supercluster LF. This is likely due to the relatively poor number counts 
for bright galaxies (which tend to be the rarer spirals, rather than the more numerous 
red sequence ellipticals). \cite{boselli2016b} remark that both Coma and A1367 contain a 
small number of bright spirals, whose star formation rates were probably increased during infall, 
while such objects are not present in Virgo. However, the composite LF in the Shapley 
supercluster \citep{haines2011} is in better agreement with Coma \citep{cortese2008} than 
with Virgo and the field studied by \citet{hammer2012}. Given the small number of clusters
observed so far, one cannot rule out other mechanisms, such as the degree of dynamical evolution,
the presence of substructure and the growth of a bright central elliptical, but our study of a
large number of clusters in this paper will allow us to examine some of these issues.

The faint end slope of the LF in Coma and A1367 by \cite{cortese2003} reaches to $M_{NUV} \sim -17$
and has a slope of $> -1.4$, while the Shapley supercluster composite LF by \cite{haines2011} has
a slope $\alpha \sim -1.5/-1.6$ to $M_{NUV}=-14$, in contrast to the Virgo cluster 
\citep{boselli2016b}, where it is flatter ($\alpha \sim -1.2$ to $M_{NUV}=-11$) and 
closer to the field value. The value in the Coma infall field of \citet{hammer2012} is intermediate
($\alpha = -1.4$ to $M_{NUV}=-10.5$) but closer to the GUVICS Virgo value. The slope of the LF
may depend somewhat on the depth reached. The steep faint end may also reflect the 
contribution from the optically bright ellipticals which are faint in the UV. For blue galaxies, 
LFs in the field and clusters are very similar, and this suggests that star formation is 
suppressed quickly upon infall \citep{cortese2008} so that only very recent newcomers to the 
cluster environment contribute to the UV LFs, while rapidly quenched galaxies are already faint 
in the NUV.

Here we consider the composite NUV LF of 28 clusters at moderate redshift ($0.02 < z < 0.08$) 
with highly complete redshift coverage of members and non members - even for  the faintest 
luminosity bins  we consider ($NUV_0\sim 20-21$ depending on cluster), mean completeness is 
above 80 per cent. We explore the effects of environment by splitting our sample into clusters 
of low and high velocity dispersion as well as low and high X-ray luminosity and sparse and 
dense clusters. Red sequence and blue cloud galaxies are also considered separately and as a 
function of cluster environment. Although our data do not reach as deep as that used in 
previous targeted studies of individual local clusters (Virgo, Coma, Abell 1367), the 
significantly larger  number of objects studied here allows us to better control stochastic effects 
from small number statistics and to explore cluster to cluster variations and therefore 
the effects of the broad cluster environment.

In the following section we describe the sample and analysis. Results are shown in section 3 and discussed in section 4. All magnitudes are in the AB system, unless otherwise noted. We assume the 
conventional cosmological parameters from the latest Planck results.

\section{Data}

Our data consist of a sample of 28 clusters at $0.02 < z < 0.08$ selected from the sample 
originally studied by \citet{depropris2003}, plus clusters from the WINGS/OmegaWINGS dataset 
\citep{cava2009,moretti2015,gullieuszik2015,moretti2017}. We selected objects with available 
GALEX imaging in the NUV (2300 \AA) band (the number of objects in the FUV band is about 1/3
of those detected in the NUV band, and we do not treat these in the present paper, although an 
analysis of the FUV data is presented elsewhere in a different context). We show the basic 
characteristics of the clusters studied in Table~\ref{table:1}. This shows the cluster 
identification, coordinates, redshift (from our work and the WINGS/OmegaWINGS dataset), velocity dispersion (also from our earlier work or as presented by WINGS/OmegaWINGS), X-ray luminosity 
(from the literature) and the IDs of the GALEX tiles used for our analysis.

\begin{table*}
   \begin{center}
   \caption{Clusters studied in this paper}
   \label{table:1}
   \begin{tabular}{@{}lcccccc}
   \hline
   Cluster & RA (2000) & Dec (2000) & $z$ & $\sigma$ (km s$^{-1}$) & $\log L_X$ (0.5--2.4 KeV) & Tiles \\
   
\hline \hline
Abell S1171 & 00 01 21.7 & $-27$ 32 18 & 0.0292 & 788 & ... & AIS\_280 \\  
Abell 2734 & 00 11 20.7 & $-28$ 51 18 & 0.0618 & 780 & 37.41 & GI1\_004001\_A2734 \\
    & & & & & & AIS\_280\\
EDCC 442 & 00 25 21.3 & $-33$ 05 23 & 0.0496 & 763 & ... & AIS\_280 \\
Abell 85 & 00 41 50.1 & $-09$ 18 07 & 0.0559 & 982 & 37.92 & GI6\_023001\_Abell85 \\
  & & & & & & GI3\_103001\_Abell85\\
Abell 119 & 00 56 18.3 & $-01$ 13 00 & 0.0442 & 503 & 37.33 & NGA\_UGC0568 \\
  & & & & & & GI1\_067001\_UGC0568 \\
    & & & & & & MISWZS\_29124\_0266 \\
    & & & & & & AIS\_265 \\
        & & & & & & NGA\_UGC0568\_css34626 \\
        & & & & & & AIS\_268 \\
            & & & & & & MISWZS01\_29158\_0269 \\
Abell 168 & 01 15 12.0 & +00 19 48 & 0.0453 & 546 & 37.53 & GI3\_121007\_J011536p002330 \\
    & & & & & & AIS\_0266 \\
    & & & & & & GI3\_121007\_J011536p002330\_css288 \\
         & & & & & & GI6\_0600006\_ESSENCE\_B2 \\
Abell S0340 & 03 20 25.5 & $-27$ 06 13 & 0.0677 & 939 & ... & AIS\_489\\
Abell 3266 & 04 31 24.1 & $-61$ 26 38 & 0.0596 & 1318 & 37.79 & GI1\_004004\_A3266 \\
    & & & & & & AIS\_420 \\
Abell 3376 & 06 01 45.7 & $-39$ 59 34 & 0.0463 & 844 & 36.81 & GI1\_004006\_A3376\\
    & & & & & & AIS\_411 \\
Abell 671 & 08 28 29.3 & +30 25 01 & 0.0502 & 956 & 36.62 & MISGCSAN\_04673\_1207\\
    & & & & & & AIS\_215 \\
Abell 930 & 10 06 54.6 & $-05$ 37 40 & 0.0549 & 907 & 35.78 & AIS\_315 \\
Abell 957 & 10 13 40.3 & $-00$ 54 52 & 0.0436 & 640 & 36.60 & AIS\_315 \\
  & & & & & & MISDR1\_24368\_0271 \\
  & & & & & & MISDR1\_24335\_0270 \\
Abell 1139 & 10 58 04.3 & +01 29 56 & 0.0398 & 503 & 37.33 & MISDR2\_12527\_0508 \\
 & & & & & & AIS\_314 \\
Abell 1238 & 11 22 58.0 & $+01$ 05 32 & 0.0733 & 586 & 36.49 & MISDR1\_12753\_0280\\
    & & & & & & AIS\_314\\
        & & & & & & MISWZN11\_12794\_0315\\
Abell 3528 & 12 45 18.2 & $-29$ 01 16 & 0.0545 & 1016 & 37.12 & GI1\_004025\_A3528 \\
    & & & & & & AIS\_491\\
    & & & & & & HRC\_RDCS1252m2927 \\
Abell 1620 & 12 49 46.1 & $-01$ 35 20 & 0.0821 & 1095 & 34.48 & MISDR1\_13592\_0.592\\
    & & & & & & AIS\_229\\
Abell 1631 & 12 53 14.4 & $-15$ 22 48 & 0.0459 & 717 & 36.86 & AIS\_330\\
Abell 3560 & 13 32 22.6 & $-33$ 08 22 & 0.0491 & 840 & 37.12 & GI1\_004012\_A3560 \\
    & & & & & & AIS\_471\\
Abell 1983 & 14 52 44.0 & +16 44 46 & 0.0436 & 522 & 36.62 & AIS\_357 \\
Abell 3716 & 20 51 16.7 & $-52$ 41 43 & 0.0457 & 848 & 37.00 & GI1\_004020\_A3716 \\
    & & & & & & GI1\_004020\_A3716\_css12332\\
        & & & & & & AIS\_364 \\
Abell 2399 & 21 57 25.8 & $-07$ 47 41 & 0.0577 & 729 & 37.27 & GI1\_004021\_A2399 \\
    & & & & & & AIS\_259 \\
        & & & & & & MISDR2\_20914\_0716\\
Abell 3880 & 22 27 52.4 & $-30$ 34 12 & 0.0548 & 855 & 37.27 & AIS\_381 \\
    & & & & & & MIS2DFSGP\_40526\_0338\\
EDCC 155 & 22 32 12.2 & $-25$ 25 22 & 0.0350 & 714 & ... & AIS\_488 \\
Abell S1043              & 22 36 26.8 & $-24$ 20 26 & 0.0340 & 1345 & ... & MIS2DFSGP\_30502\_0066 \\
  & & & & & & AIS\_488 \\
Abell 2660 & 23 45 18.0 & $-25$ 58 20 & 0.0525 & 845 & 35.70 & AIS\_279\\
Abell 4038 & 23 47 43.2 & $-28$ 08 29 & 0.0282 & 933 & 37.10 & AIS\_279 \\
Abell 4053 & 23 54 46.7 & $-27$ 40 18 & 0.0720 & 994 & ... & AIS\_280 \\
Abell 4059 & 23 57 02.3 & $-34$ 45 38 & 0.0490 & 752 & 37.49 & GI1\_004023\_A4059 \\
    & & & & & & AIS\_280 \\
 \hline
\hline
   \end{tabular}
   \end{center}  
\end{table*}

For each cluster we retrieved the appropriate tile(s) and used the available (Kron-style) 
photometry \citep{martin2005,morrissey2007} in the NUV to select objects. We only considered 
detections within each cluster's $r_{200}$ (calculated from the formula in \citeauthor{carlberg1997} 
\citeyear{carlberg1997}) and with a S/N of at least 5 in the NUV band.  NUV magnitudes were 
corrected for extinction using the values provided in NED \citep{schlafly2011}. However, for 
some clusters $r_{200}$ exceeded the size of the GALEX tiles and we only consider objects in the 
inner $0.5^{\circ}$ of each tile, where GALEX has full sensitivity. Our spatial coverage is 
comparable or greater than  that obtained in prior targeted studies of  nearby clusters such 
as Virgo, Coma and Abell 1367. 

GALEX has a resolution of about $5''$ and it is therefore not possible to carry out star-galaxy 
separation from the data. For this purpose we cross-matched each of the GALEX detections with 
CCD images from the PanStarrs1 survey, which covers the entire sky north of $\delta=-30^{\circ}$
\citep{chambers2016} and also provides optical photometry \citep{magnier2016,flewelling2016}, 
except for some clusters in the South where we used photographic data from the UK Schmidt 
Telescope SuperCosmos survey \citep{hambly2001a,hambly2001b}. For each $NUV$ detected object
we assign a star/galaxy type based on $r_{PSF}-r_{Kron} < 0.05$ mag. (for a star) and direct
inspection of each image with {\tt imexamine} if necessary. For clusters outside of the PanStarrs
footprint we instead use the star/galaxy class provided by the SuperCosmos survey, and we also check
these objects visually. The matching radius was $6''$, equivalent to the NUV PSF and similar to 
that used in previous studies. The optical photometry is also corrected for Galactic extinction from
NED. We chose very conservative NUV magnitude limits, such that all our galaxies have optical 
counterparts, although this of course means that our LFs do not reach as faint as other work in 
more nearby clusters. In any case, the limiting magnitude of our study is set by the much 
brighter spectroscopic completeness limit, as detailed below.

We then matched (also within $6''$) our galaxies with redshifts within NED (NASA extragalactic 
database) and other sources (such as the WINGS/OmegaWINGS catalogues in \citeauthor{cava2009} 
\citeyear{cava2009}, \citeauthor{gullieuszik2015} \citeyear{gullieuszik2015}, \citeauthor{moretti2017}
\citeyear{moretti2017}). Most of the redshifts come from the 2dF and SDSS surveys as well as the 
WINGS/OmegaWINGS survey, but a few other sources of measurements are occasionally indicated in NED.

The redshift information used comes from a variety of sources, some of which may suffer from 
selection effects, especially as the redshift catalogs are selected in the optical while our 
photometric catalog is selected in the NUV. Our analysis is based on the methods originally 
developed in \cite{depropris2003} and \cite{depropris2017}, where we correct for incompleteness 
based on the redshift distribution (as a function of magnitude) for field and cluster galaxies 
(also see \citeauthor{mobasher2003} \citeyear{mobasher2003}). In order to do this, we need to 
first make sure that the redshift information is not biased towards or against cluster members, as 
this lack of bias is an assumption of the method.

\begin{figure*}
\includegraphics[width=\textwidth]{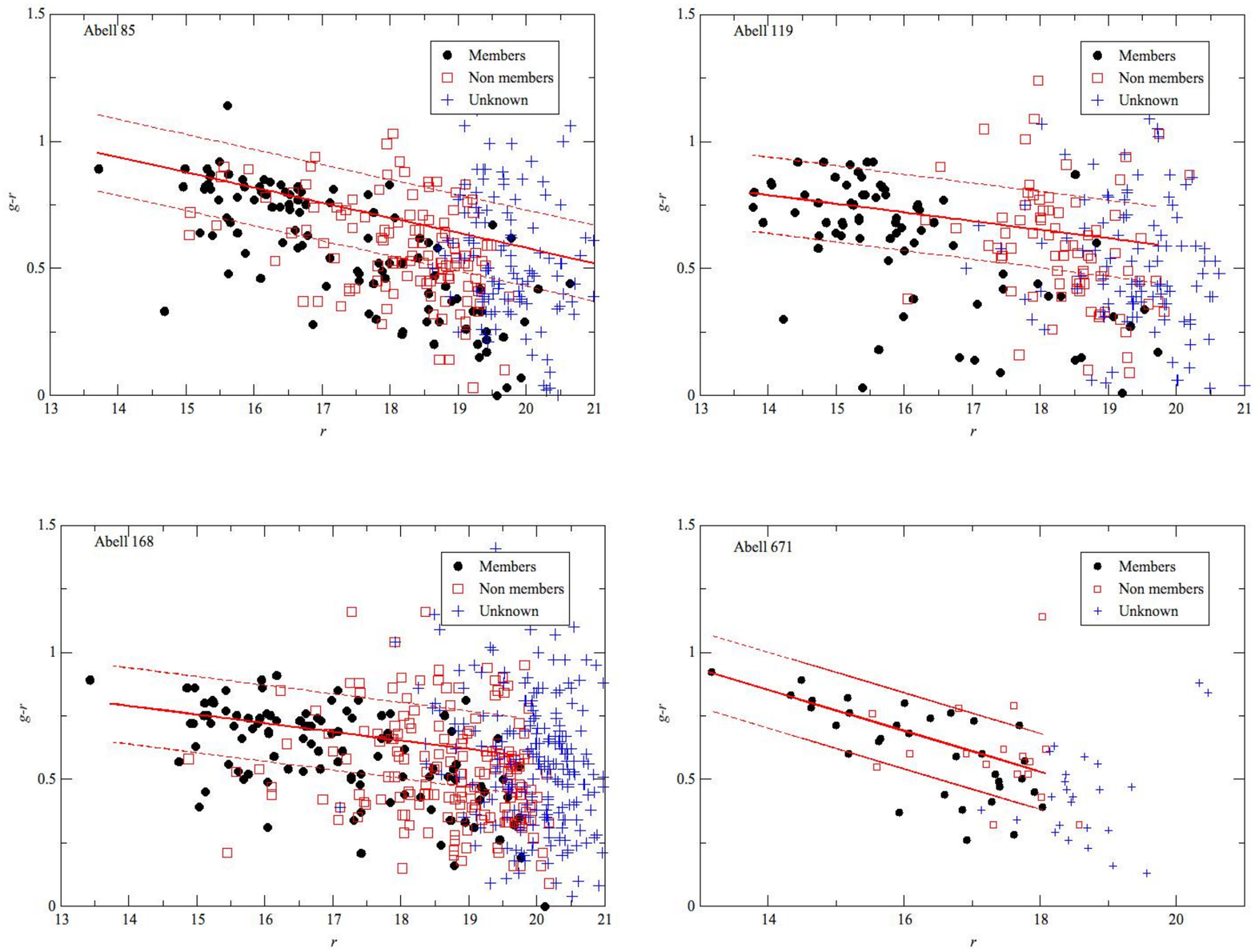}
\caption{Colour-magnitude diagrams in $g-r$ (or $B_J-R_F$) vs. $r$ for NUV-selected galaxies 
in our clusters. Symbols as identified in the figure legends. Here we show a representative
sample of objects. All remaining clusters are shown in the on-line appendix.}
\label{fig:1}
\end{figure*}

As a first step, we therefore plot the $g-r$ vs. $r$ colour-magnitude diagrams for all NUV-selected 
galaxies in Fig.~\ref{fig:1}, and fit the red sequence for each cluster using a minimum 
absolute deviation least squares fitting procedure \citep{beers1990} to remove the influence of 
outliers. Here we show only the first four clusters. The 
remaining clusters can be seen in the on-line appendix (with the same format). For clusters below
$-30^{\circ}$ in declination, that are not in the PanStarrs footprint, we use the Supercosmos 
$B_J-R_F$ and $R_F$ colours instead. Note that some clusters do not have a clear colour-magnitude
relation with our data, and these are not plotted in Fig.~\ref{fig:1}. Our previous work has shown 
that the colour dispersion about the red sequence is 0.05 mag. in $g-r$ \citep{depropris2017}, in 
agreement with other studies (e.g., \citeauthor{valentinuzzi2011} \citeyear{valentinuzzi2011}). We
therefore exclude all objects (without a redshift) 0.15 mag. redder than the red sequence, 
as these are unlikely to be members. We can see from Fig.~\ref{fig:1} that the majority of objects
redder than the red sequence for which a redshift is available are non-members and the remaining
galaxies with unknown redshift therefore have the same colour distribution as the sample of galaxies
with known redshifts (irrespective of their being cluster members); these former objects tend to
lie at the faint end of the spectroscopic sample.

\begin{figure*}
\includegraphics[width=\textwidth]{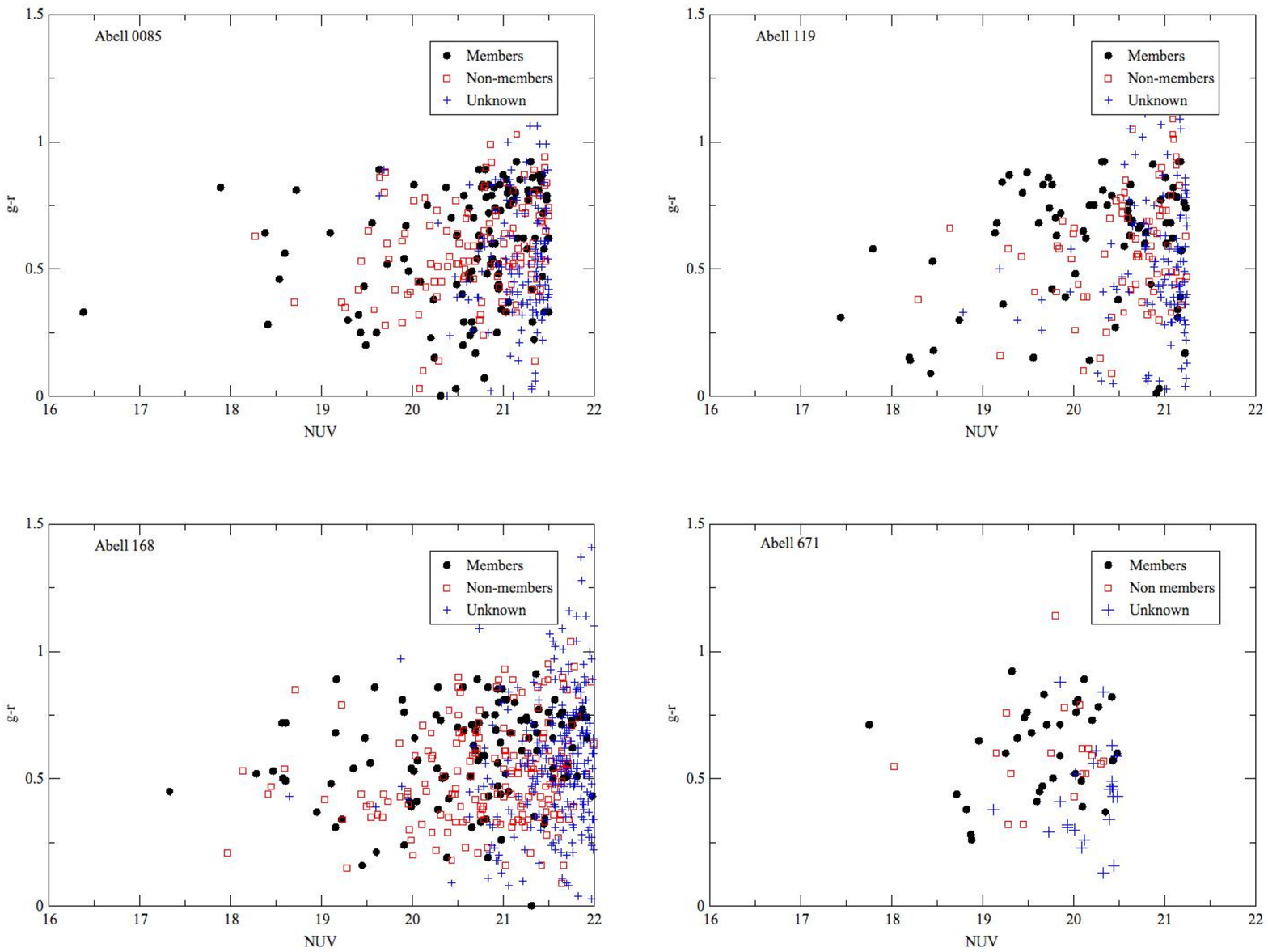}
\caption{Same as Fig.~\ref{fig:1} but for $g-r$ vs. NUV, with the same symbols (as indicated in the figure legend). This shows that NUV bright cluster members tend to be optically blue while the massive red sequence galaxies are generally faint in NUV. All remaining clusters are shown in the on-line appendix.}
\label{fig:2}
\end{figure*}

In Fig.~\ref{fig:2} we plot the equivalent of Fig.~\ref{fig:1} but for $NUV_0$ (i.e., the band
we select galaxies in). While red sequence galaxies can be easily identified even when plotting
$g-r$ vs. $NUV_0$, it is apparent that the optically bright red sequence galaxies are faint in 
$NUV$ whereas the brighter galaxies tend to be relatively optically faint systems. For blue 
galaxies, NUV flux comes primarily from young stars, whereas for red galaxies UV flux is primarily 
provided by hot horizontal branch stars  that produce the ultraviolet excess \citep{schombert2016}.
We elaborate on this further in this paper.

\begin{figure*}
\includegraphics[width=\textwidth]{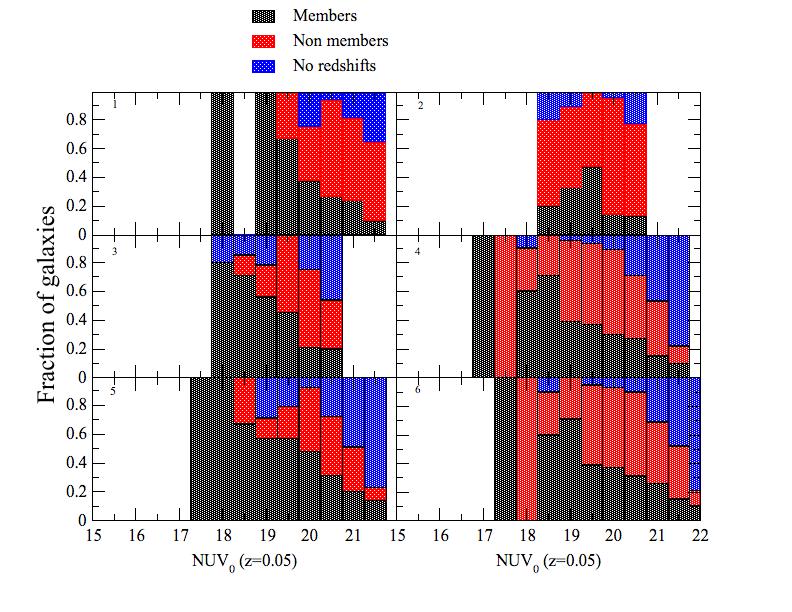}
\caption{Completeness fractions for each cluster vs NUV$_0$ magnitude (in the $z=0.05$ bins described in the text). The numbers in the top left corner of each panel identify the cluster, in the order shown in Table 1 (here 1: Abell S1171; 2: Abell 2734; 3: EDCC 442; 4: Abell 85; 5: Abell 119; 6: Abell 168). The symbols used in the plot are identified in the figure legend. All remaining clusters are shown in the on-line appendix.}
\label{fig:3}
\end{figure*}

In Fig~\ref{fig:3} we show the redshift completeness for all clusters as a function of $NUV$ magnitude.

\section{The Luminosity Functions}

We produce composite UV LFs of UV detected galaxies in the same fashion as in our previous studies of cluster optical and near-IR LFs \citep{depropris2003,depropris2017}. We count galaxies in each cluster in 0.5 mag. bins shifted to their corresponding values at $z=0.05$, which is the mean redshift of the sample in Table~\ref{table:1}. For each cluster we calculate the difference in distance modulus between its redshift and $z=0.05$. We then count cluster members in apparent 
magnitude bins corresponding to the fixed magnitude bins at $z=0.05$. For instance, in Abell 1139 ($cz=11711$ km s$^{-1}$) the magnitude 
interval $14.81 < NUV_0 < 15.31$ contributes to the galaxy counts in the $15.25 < NUV_0 < 15.75$ bin at $z=0.05$, whereas in Abell 2734 
($cz=18737$ km s$^{-1}$) counts in the apparent magnitude bin $15.66 < NUV_0 < 16.16$ contribute to the equivalent $15.25 < NUV_0 < 15.75$ 
counts at $z=0.05$. 

In this way we avoid carrying out uncertain (and variable from galaxy  to galaxy) $e$ and $k$ corrections to $z=0$. Since only a small shift of $0.03$ (maximum) in redshift is needed, the relative $e+k$ corrections will be 
small. The modest error induced for clusters at $z < 0.05$ is compensated by an error (of the same magnitude but opposite sign) for 
clusters at $z>0.05$ so that the composite LFs at $z=0.05$ should not suffer from these uncertainties. A similar approach is used by 
\citet{blanton2003} and \citet{rudnick2009}. For the assumed cosmology, the distance modulus to $z=0.05$ is 36.64 mag. but this does not include the effects of $e+k$ corrections.

In order to create a composite LF we need to correct for incompleteness. The procedure is analogous to that used in our previous work \citep{depropris2003,depropris2017}. Fig.~\ref{fig:2} shows the completeness fractions as a function of 
observed $NUV_0$ magnitude for all our clusters, including members, non members and objects for which no redshift is known. We then correct 
for incompleteness in the same manner as in \citet{depropris2003} and \citet{mobasher2003}. We then calculate the composite LF following the procedure detailed in \cite{colless1989} where we generally stop the summations (for each cluster) at completeness levels of 80\% or more. 

Here the number of members for cluster $i$ in each magnitude bin $j$ is given by:

\begin{equation}
N_{ij}=\frac{N_I N_M}{N_Z} 
\end{equation}

where $N_I$ is the number of galaxies in the input photometric catalog, $N_M$ the number of cluster members and $N_Z$ the number of objects with a measured redshift (whether a cluster member or not) in each bin.
 

Here $N_I$ is a Poisson random variable whereas $N_M / N_Z$ is a binomial variable. The mean and variance for the approximately normal distribution of the sample proportion are $p$ and $(p(1-p)/n)$, where $p$ is the probability of a `success'(i.e. a cluster member) $N_M/N_Z$ and $n$ is the number of trials $N_Z$. Therefore the error in this quantity can be written as (ignoring a
small negative second order term):


\begin{equation}
\frac{\sigma^2 (N_{ij})}{N_{ij}^2} = \frac{1}{N_I} + \frac{1}{N_M}-\frac{1}{N_Z}
\end{equation}

We then calculate the composite LF following the procedure detailed in \cite{colless1989} where we generally stop the summations (for each cluster) at completeness levels of 80\% or more. 

\begin{equation}
N_{cj}={N_{c0} \over m_j} \sum_i {N_{ij}\over N_{i0}}
\end{equation}

where $N_{cj}$ is the number of cluster galaxies in magnitude bin $j$, and the sum is carried
over the $i$ clusters and $m_j$ is the number of clusters contributing to magnitude bin $j$. Here
$N_{i0}$ is a normalisation factor, corresponding to the (completeness corrected) number of galaxies 
brighter than a given magnitude (here we use $NUV_0=19.0$) in each cluster and

\begin{equation}
N_{c0}=\sum_i N_{i0}
\end{equation}

The error is then given by:

\begin{equation}
\delta N_{cj}={N_{c0} \over m_j} \Bigg[ \sum_i \Bigg({\delta N_{ij} \over N_{i0}}\Bigg)^2\Bigg]^{1/2}
\end{equation}

{\bf These summations are carried out individually and separately for each of the subsamples 
discussed below, so that the total numbers do not always scale between the various LFs.}

\subsection{Luminosity functions for all galaxies}

\begin{figure}
\includegraphics[width=0.45\textwidth]{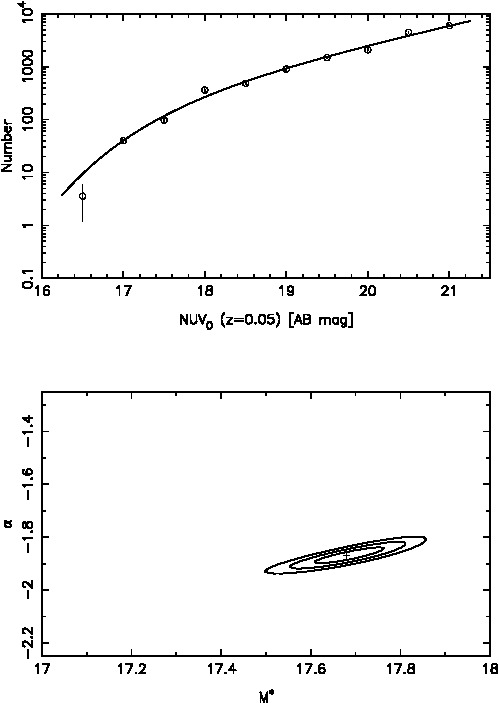}
\caption{Luminosity function and best fit for all clusters in the sample (top). 
Error ellipses for 1, 2 and 3$\sigma$ are shown (bottom). In this and subsequent figures 
$M^*$ refers to the observed characteristic magnitude of the LF measured at $z=0.05$.}
\label{fig:4}
\end{figure}

We fit the composite NUV LF for all 28 clusters with a standard Schechter function using a 
Levenberg-Marquardt minimisation algorithm. This is shown in Fig.~\ref{fig:4} together with the 
associated error ellipses (the LF parameters are strongly correlated). The values for $M^*$ and 
$\alpha$ are tabulated in Table~\ref{tab:2} where we also present conditional 1$\sigma$ errors 
for each parameter (i.e., the error holding the other parameter fixed).

For all clusters the best fitting LF has $M^*_{NUV}=-18.98 \pm 0.07$ and $\alpha=-1.87 \pm 0.03$ 
(assuming the above distance modulus and no k+e correction, as noted earlier). The value we derive 
for this $M^*$ is in good agreement with that measured by \cite{cortese2008} for the Coma cluster,
fainter (but consistent at the $\sim 1.5\sigma$ level) than the one measured by \cite{cortese2005} 
in Abell 1367 and considerably brighter than the GUVICS LF in the Virgo cluster \citep{boselli2016b}. 
It is however in very good agreement with $M^*$ for the composite LF of the 5 Shapley supercluster 
clusters in \citet{haines2011}. The LF we derive also has a somewhat brighter $M^*$ than the local 
field in \cite{wyder2005}, although this latter measurement has large uncertainties (it only differs 
at the 1.5$\sigma$ level). At face value this might indicate a brightening of the typical UV flux 
in cluster galaxies, possibly due to induced star formation upon first encountering the cluster 
enviroment. The slope  is quite steep, $\alpha \sim -1.9$. This is in broad agreement with 
the values for $\alpha$ obtained for Coma, Abell 1367 and the Shapley clusters, but not with the 
much flatter slopes derived for the field and Virgo as well as the region in the Coma cluster studied 
by \citet{hammer2012}: these latter have $\alpha \sim -1.2$  to $-1.4$. 

The differences in the bright end of the LF may be more naturally explained by small number 
statistics. Unlike optical LFs, the bright end of UV LFs are dominated by star-forming spirals 
(see, e.g., Fig.~\ref{fig:1}). These objects are likely to be a transient and comparatively 
stochastic population that has recently become bound to the cluster \citep[e.g.][]
{cortese2008,haines2011}. It is therefore understandable that local conditions may drive 
their contribution to the bright end of the LF. Our composite LF is less sensitive to this, 
as it averages contributions from several clusters, spanning a wide range of properties. It 
is perhaps indicative that our $M^*$ is in good agreement with the composite LF of \cite{haines2011}.

The faint end of the NUV LF includes components from fainter star-forming galaxies as well as 
bright early-type galaxies. As these objects tend to be brighter in the UV as a function of 
optical luminosity (e.g., \citeauthor{Ali2018} \citeyear{Ali2018} ), this leads to steeper NUV LFs in clusters (bright early-type galaxies 
are much rarer in the field), see below for our discussion on this. On the other hand, the LF 
slopes are much flatter in Virgo \citep{boselli2016b} and the in outskirts of Coma \citep{hammer2012}. 
These authors suggest that uncertainties in completeness corrections and differences in depth 
(between their work and Coma/Shapley) are responsible for this effect. While our data are not as 
deep as those in nearby clusters, they reach $\sim 2.5$ mag. below the $L^*$ point, where 
$\alpha$ should be well determined. Our completeness corrections are well understood and small 
(as discussed above). This may point to real differences in the faint end slope between our 
clusters and the regions studied by GUVICS and  \cite{hammer2012}, likely due to the behaviour of 
faint star-forming galaxies in these different environments (to which these objects 
should be especially sensitive). This is also further discussed in the next sections.

\begin{table}
   \begin{center}
   \caption{Schechter function parameters when fitted to full available $NUV_0$ range for each subsample. Errors are statistical, and consequently do not always reflect whether  a Schecter function is an appropriate fit to the data. }
   \label{tab:2}
   \begin{tabular}{@{}lcc}
   \hline
   Sample & $m^*_{NUV_0}\  (z=0.05)$ & $\alpha$\\
\hline \hline
All clusters & $ 17.66 \pm 0.07$ & $-1.87 \pm 0.03 $ \\
All clusters (red galaxies) & $17.42 \pm 0.28$ & $-2.14 \pm 0.07$\\
All clusters (blue galaxies) & $18.02 \pm 0.09$ & $-1.55 \pm 0.04$ \\
$\sigma < 800$ km s$^{-1}$ all galaxies & $18.24 \pm 0.08$ & $-1.57 \pm 0.04$ \\
$\sigma > 800$ km s$^{-1}$ all galaxies & $18.21 \pm 0.09$ & $-1.59 \pm 0.04$ \\
Low $L_X$ all galaxies & $ 17.91 \pm 0.17$ & $-1.84 \pm 0.09$ \\
High $L_X$ all galaxies & $19.84 \pm 0.06$ & $-1.13 \pm 0.07$ \\
Dense clusters all galaxies & $ 17.33 \pm 0.18$ & $-1.91 \pm 0.04$ \\
Sparse clusters all galaxies & $ 18.40 \pm 0.09$ & $-1.55 \pm 0.09$ \\
 \hline
\hline
   \end{tabular}
   \end{center}  
\end{table}

In order to address the above issues, we now consider the LFs of optically red and blue galaxies 
separately. In previous studies star-forming galaxies and quiescent galaxies were separated 
according to their NUV-$r$ colour. Given the strong sensitivity of the NUV colour to star formation,
this sharply separates star-forming and quenched galaxies. However, this may lead to a selection 
effect, as shown by \citet{cortese2008}, where recently quenched galaxies cluster on the red 
sequence and only currently star-forming galaxies are observed in the `blue cloud'. Because star 
formation is expected to be strongly suppressed in clusters, these galaxies represent a transient 
population of objects newly accreted from the field. This means that one observes the same blue 
galaxy LF in all environments, because all sensitivity to environmental effects is masked by
the rapid quenching process and the sensitivity of the NUV band to star formation, although in 
reality the process of colour and morphological evolution is likely to be more prolonged 
\citep{bremer2018}.

Here we instead select blue and red galaxies from the optical $g-r$ colours in Fig.~\ref{fig:1}.
Red sequence galaxies lie within $\pm 0.15$ mag. (3 $\sigma$ given the measured colour 
dispersion) of the best fitting straight line shown in Fig.~\ref{fig:1}, while our blue cloud 
galaxies are all objects bluer than the blue edge of the red sequence as defined above.
Because the $g-r$ colour changes more slowly after star formation declines, our choice separates
galaxies into long term quiescent objects (with possibly a small fraction of galaxies showing 
residual star formation) and star-forming objects plus ``green valley'' galaxies. Our choice allows 
us to explore the effects of the cluster environment on star-forming galaxies on a longer timescale,
closer to the crossing time of clusters and comparable to the timescale over which galaxies cross 
the green valley in \citet{bremer2018}. 

\begin{figure*}
\includegraphics[width=0.36\textwidth,angle=270]{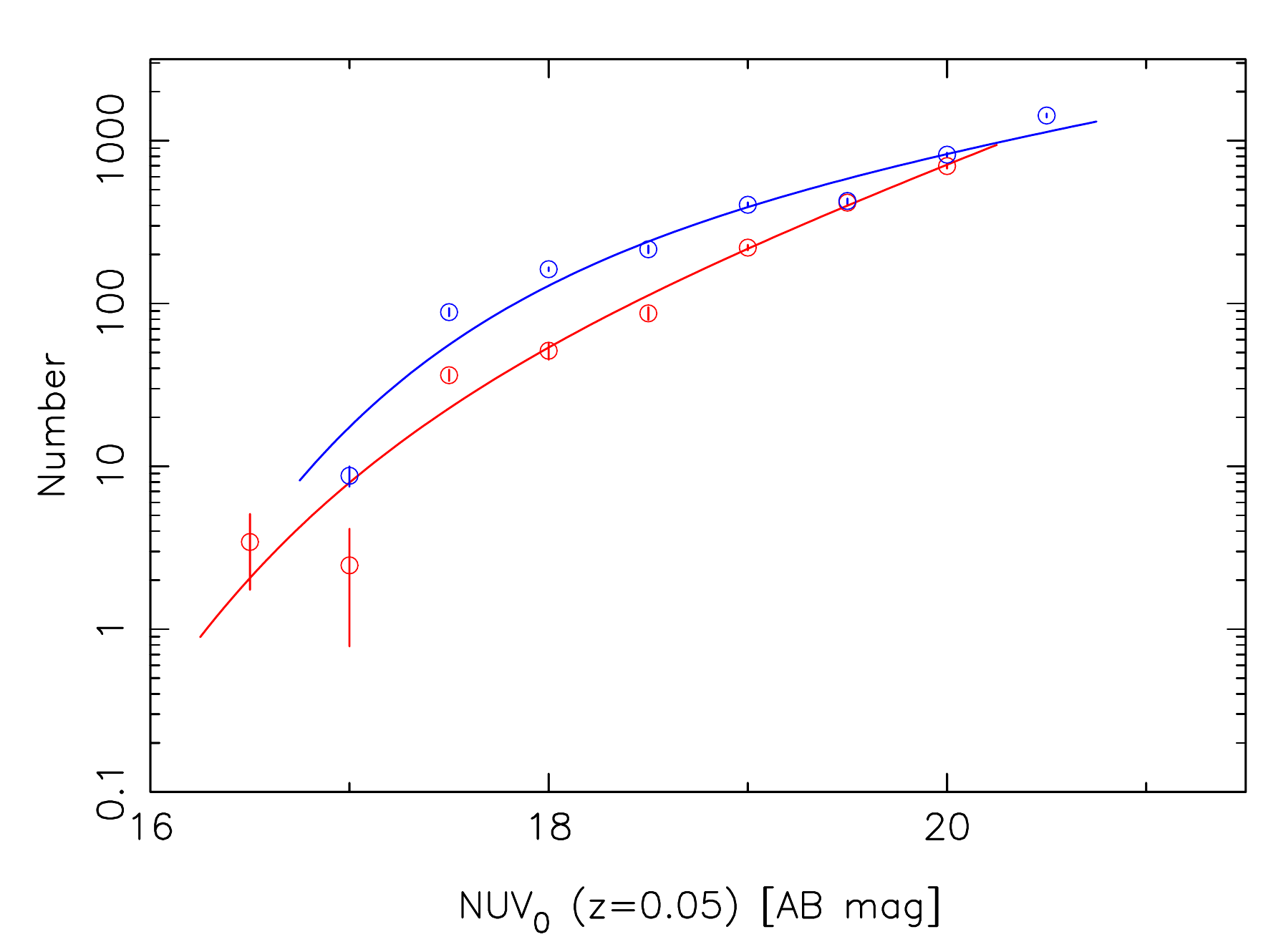}
\includegraphics[width=0.36\textwidth,angle=270]{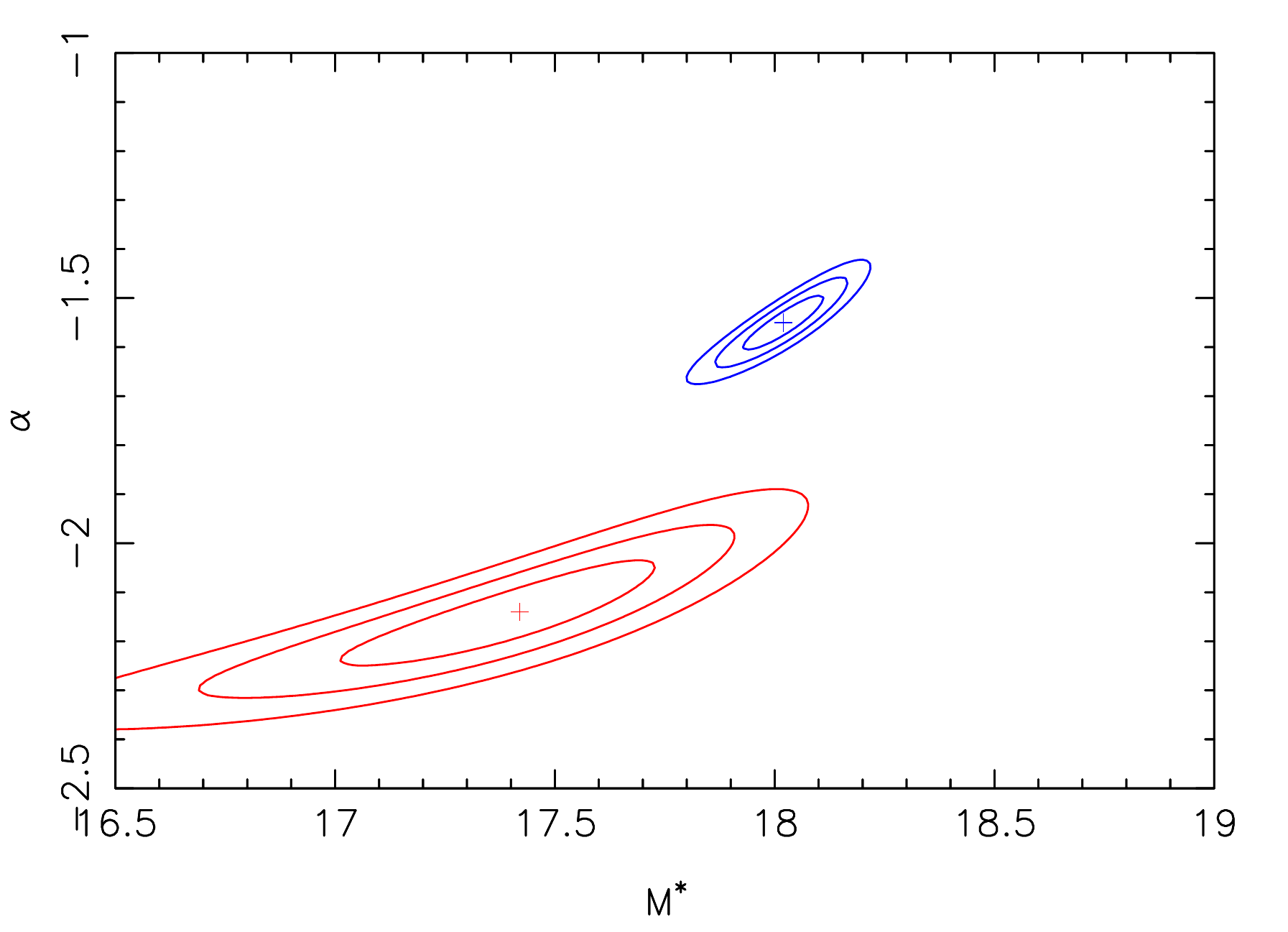}
\caption{Luminosity function and best fit for red and blue galaxies (identified by colour) for all clusters in the sample (left). Error ellipses for 1, 2 and 3$\sigma$ are shown in the right-hand panel, for red and blue galaxies}
\label{fig:5}
\end{figure*}

\begin{figure*}
\includegraphics[width=\textwidth]{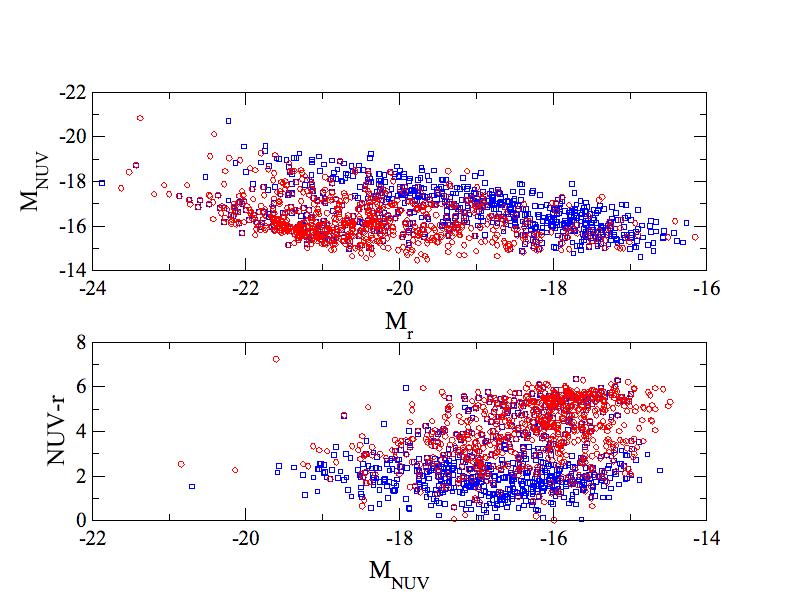}
\caption{For all galaxies in all clusters we plot the absolute magnitudes in $r$ vs.
the absolute magnitudes in $NUV$ on the top panel. Red symbols are for red sequence galaxies and blue
symbols for blue cloud galaxies. On the bottom panel we show the $NUV-r$ colours as a function of absolute NUV magnitude.}
\label{fig:6}
\end{figure*}

We plot the LFs for red and blue galaxies in Fig~\ref{fig:5}, with the parameters tabulated in
Table~\ref{tab:2}. The Schechter function is not a very good fit to the red sequence LF,
however the slope is very steep. Fig~\ref{fig:6} plots the absolute $M_r$ vs. absolute $M_{NUV}$ for all 
galaxies, with different symbols for the optically red and blue samples. This shows that 
(a) red galaxies are faint in $NUV$ and contribute mainly to the faint end of the NUV LFs,
giving rise to the steep slopes (see upper panel), (b) the red galaxies are nevertheless 
optically bright and therefore generally more massive, being typically $\sim L^*$ or brighter 
in the optical (also see upper panel in this figure),  whereas (c) blue galaxies are bright in NUV
, but are usually optically faint (upper and lower panels of Fig~\ref{fig:6}). Star forming
galaxies in low redshift clusters are generally sub-$L^*$ systems. Fig.~\ref{fig:6} (upper panel)
shows how optically bright galaxies (i.e., $-20 < M_r < -22$) mainly lie at $ -15 < M_{NUV} < -17$ 
i.e., they are usually 1--2 magnitudes fainter than the $M^*$ point for the NUV LF. 
Fainter red sequence galaxies are missing from our sample unless they are star-forming and have 
bluer $NUV-r$ colours because of a selection effect. This is because the $NUV$-optical colour 
for quiescent galaxies becomes bluer with increasing luminosity as the NUV emission is produced 
by hot horizontal branch stars (rather than star-formation), so that we are looking at the 
exponential part of the galaxy mass function, as seen in the lower panel of Fig.~\ref{fig:6}. 
Therefore the optically bright galaxies mainly contribute to the faint end of the NUV LF 
and this results in a steep total LF slope. Unfortunately, no previous work seems to treat the 
red galaxies LF in detail, except for the study by \citet{boselli2016b} in the Coma cluster, 
compared to which our $M^*$ is somewhat brighter and our $\alpha$ is steeper.

\begin{figure*}[h!]
\centering
\includegraphics[width=0.9\textwidth]{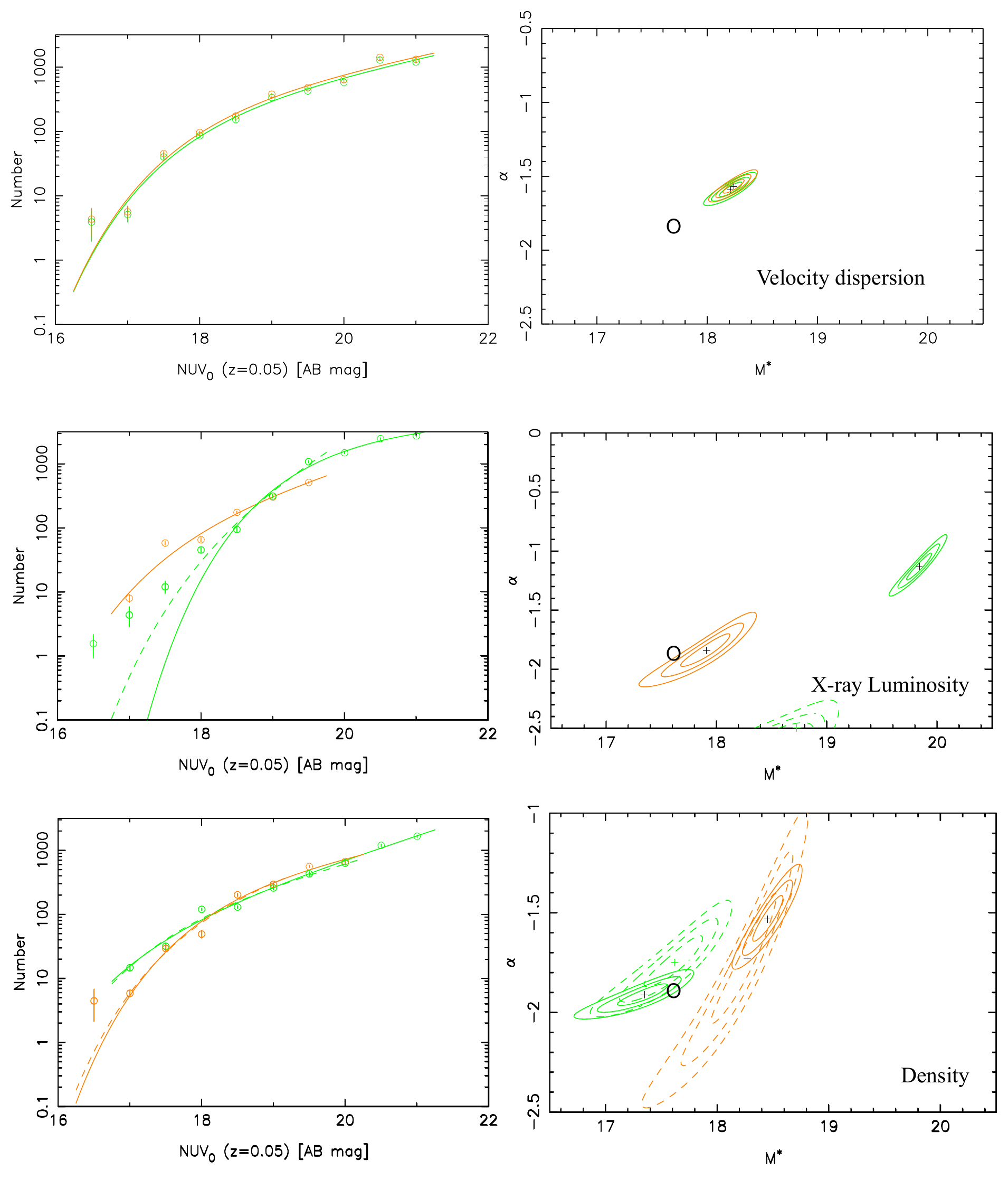}
\caption{Luminosity functions and
best fits (left column) and associated error ellipse (right column) for low (orange) and high (green) $\sigma$ clusters (top row), low (orange) and high (green) $L_X$ clusters (middle row) and sparse (orange) and dense (green) clusters (bottom row).
{\bf The circle in each figure indicates the $M^*$ and $\alpha$ values for the total LF.} Clearly, for the subsamples split by X-ray luminosity and surface density, the luminosity distributions as sampled here are not well fitted by Schechter functions, the location and size of the error ellipses in these cases should be used with caution. Any interpretation should take note of the actual numbers rather than the fits to the data. The continuous lines and contours are the fits to the entire range of $NUV_0$ values available for a given subsample. The dashed lines and contours are for Schechter fits restricted in range in $NUV_0$ where both subsamples have data. Clearly, in the case of splitting by $L_x$, even this does not result in a good fit for the high $L_x$ subsample, though obviously the data imply  a lower ratio of high to low $L_{UV}$ galaxies in the high $L_x$ subsample relative to that of the low $L_x$ subsample. For the split by surface density, restricting the fitting range has no real effect.}
\label{fig:7}
\end{figure*}

For the UV luminosity function of blue galaxies, we obtain (for the above stated distance modulus) 
$M^*=-18.62 \pm 0.08$ and $\alpha=-1.55 \pm 0.04$. These values are in good agreement with the 
values for star-forming galaxies in \cite{haines2011} but both brighter and steeper than the 
GUVICS LF for Virgo \citep{boselli2016b} and the Coma LFs of \cite{hammer2012}. The blue LF we 
derive has a similar $L^*$ to the local field,  but a somewhat steeper slope \citep{wyder2005}. 
This would suggest that star-forming galaxies are not strongly affected by the cluster environment 
or represent a transient population of objects that have not yet been quenched, as in 
\citet{cortese2008}. The somewhat steeper slope than in the field may indicate the presence of 
star-forming low mass galaxies, although the errors on the field LF are too large for this conclusion to be secure.

\subsection{Environmental effects on the total LFs}

We can now consider the effects of environment by splitting our clusters into sub-samples. 
We first divide our sample according to velocity dispersion, a broad indicator of cluster mass. 
We use $\sigma=800$ km s$^{-1}$ to identify `low mass' and `high mass' clusters as in 
\cite{depropris2017}. Fig.~\ref{fig:7} shows the derived LFs for both subsamples, with the 
best fitting parameters given in Table~\ref{tab:3}. The derived $m^*$ (at $z=0.05$) and 
$\alpha$ are consistent with each other and with the total LF shown in Fig.~\ref{fig:4}. 
This suggests that cluster velocity dispersion (or cluster mass) does not strongly affect 
the NUV LF, or that these measured velocity dispersions are not completely reliable as a 
proxies for mass. If we assume that differences in environment will only affect the blue 
population, as the NUV flux from red galaxies comes from old stellar populations, rather 
than star formation, this suggests that environmental effects are not strong or that blue 
galaxies are a transient population similar to field galaxies and rapidly quenched. Of course 
this implies that the red galaxy LF should be more strongly affected as it includes recently 
quenched objects, but these may have low masses \citep{cortese2008}, and therefore UV luminosities 
too low for them to be included in our UV-selected sample. The differences reported between 
Virgo, Coma, A1367 and the Shapley supercluster may be due to stochastic variations, especially 
at the higher luminosity end, rather than physical effects. The small number of clusters
observed in previous work and the variety in the fields observed (e.g., in Coma the central
region containing most of the bright early-type population has not been included in the 
study by \citeauthor{hammer2012}\citeyear{hammer2012}) means that other effects such as
dynamical status or Bautz-Morgan type could not be addressed.

We also consider clusters with high and low X-ray luminosity measured in the 0.5--2.4 KeV band. 
The X-ray luminosities are taken from the BAX database \citep{sadat2004} and come from a variety 
of sources, although most are derived from \cite{bohringer2017} and \cite{piffaretti2011}. 
We split the clusters into high and low X-ray luminosity subsamples at $L^*_X= \sim 1.2 \
times 10^{37}$ W \citep{bohringer2014}. Not all clusters have a measured X-ray flux,  
as noted in Table~\ref{table:1}. We show the LFs for the low and high X-ray luminosity subsamples 
in Fig.~\ref{fig:5} with values for their Schechter  parameters tabulated in Table~\ref{tab:2}. 
The Schechter function is, however, a poor fit to the high X-ray luminosity cluster LF, with a 
few bright galaxies at $NUV < 17$ affecting the fit. The LF for low X-ray luminosity clusters 
is broadly consistent with the total LF, though still not a good fit. Regardless of how well 
these are fitted by Schechter functions, there are clear differences between the two LFs. 
Ignoring the fits and simply looking at the relative numbers at the bright and faint end of 
the two LFs (i.e. their overall slope), the high X-ray luminosity distribution has relatively 
fewer high UV luminosity galaxies, suggesting  that star formation has been suppressed in 
these clusters.  The observations therefore may suggest that interactions with the X-ray gas, 
such as ram stripping, play an important role in quenching star formation, even  in massive 
cluster galaxies.

Finally, we divide our sample into dense and sparse clusters, based on the surface density 
of red sequence galaxies (tracing the cluster mass) brighter than $M_r=-20$. The resulting 
LFs are shown in Fig.~\ref{fig:5} as well and tabulated in Table~\ref{tab:2}. Here we see that 
sparse clusters more closely resemble the field values, while dense clusters are closer to the 
value for the total galaxy LF in Fig~\ref{fig:4}. In dense clusters the NUV LF is both brighter
(in terms of $ M^*$) and steeper. We show below that this reflects the contribution from red 
sequence ellipticals rather than star-forming galaxies. These objects are rather rarer in the 
sparser clusters. Indeed, the sparse cluster subsample lacks any galaxies fainter than 
$NUV_0=20$ in figure \ref{fig:7}.

\subsection{Environmental effects on red and blue galaxies}

In these subsamples we can now consider blue and red galaxies (as defined above), separately.
Unfortunately, the NUV LFs for red galaxies only cover a few magnitudes and are generally
poorly fitted. Our subsamples contain too few galaxies for our procedure to return a good
fitting LF for the red population. Since these objects are quiescent, there should be no 
significant effects from the environment on their NUV luminosities, where the flux is expected 
to come from hot horizontal branch stars or residual star formation in already quenched galaxies.

\begin{figure*}
\includegraphics[width=\textwidth]{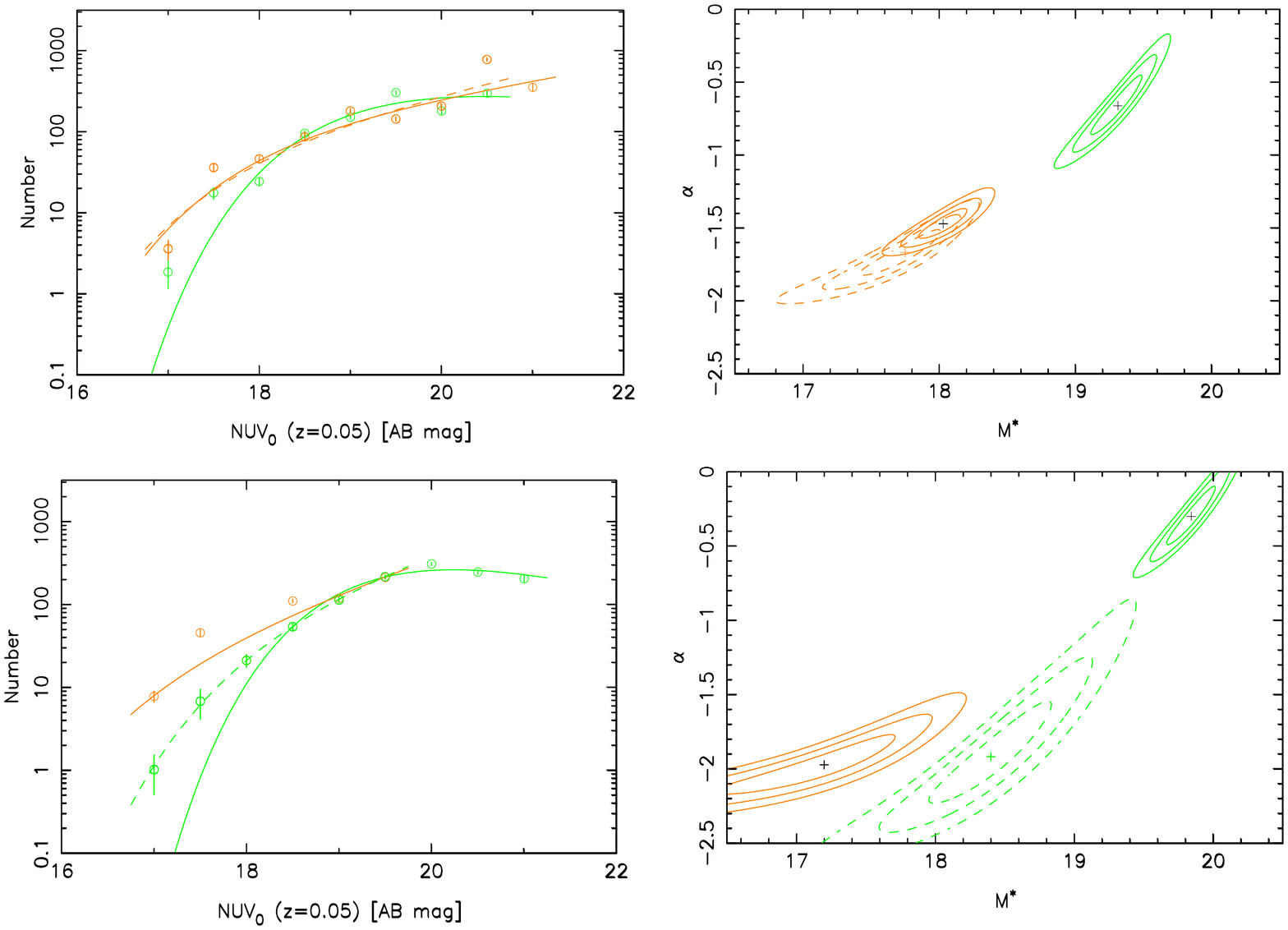}
\caption{Left columns: Luminosity functions for blue galaxies in high (green) and low $\sigma$ (orange) clusters (top) and for low $L_X$ (orange) and high $L_X$ (green) clusters (bottom). Right columns: the corresponding error ellipses. Again, care should be taken over the interpretation of the fitted parameters given the quality of the fits, though it is clear when simply looking at the data points that at least in the case of splitting by X-ray luminosity, there is a clear difference by the two subsamples.Continous and dashed lines and contours have the same meaning as in figure 6. The dashed line and contours for the split by X-ray luminosity show the (improved) fit for the high $L_x$ subsample when only ranging over the same $NUV_0$ values as for the low $L_x$ subsample. In this case, the faint end slopes between the two subsamples are similar and the main difference is in a $\sim 1$~mag shift in $M^*$, it being fainter for the high $L_x$ subsample.}
\label{fig:8}
\end{figure*}

We therefore only derive LFs for blue galaxies in all our subsamples of clusters. The resulting
LFs and error ellipses are shown in Fig.~\ref{fig:8} while we tabulate the values for the
best fitting parameters in Table~\ref{tab:3}. It is clear that the LFs of the blue galaxy 
populations show significant differences as a function of environment. The blue galaxies in 
low velocity dispersion clusters have a  LF closer to that of the  field, while in high 
velocity dispersion clusters the LF is both fainter (in terms of $M^*$) and has flatter slope. 
This is again consistent with a decrease in star formation rate and a deficit of low UV 
luminosity galaxies in the more massive clusters. The same effect is seen, possibly at a stronger 
level, in the case of splitting the sample by $L_x$, although the Schechter function is a poor 
fit to the LF of the high $L_x$ subsample, even when restricted in fitting range. Nevertheless 
the difference in the ratio of UV bright to faint galaxies between the subsamples is clear from 
the data, regardless of the fit. Unfortunately, we cannot reliably compare LFs for blue galaxy 
subsamples split by surface density as it is only possible to determine a LF for blue galaxies 
in the dense clusters. For consistency, we do not show the blue galaxy LF for dense clusters, 
though we note that this is essentially the same as the LF for {\it all} of the blue galaxies.

\begin{table}
   \begin{center}
   \caption{Schechter function parameters for blue galaxies when fitted over full available range of $NUV_0$ for each subsample}
   \label{tab:3}
   \begin{tabular}{@{}lcc}
   \hline
   Sample & $m^*_{NUV_0}\ (z=0.05)$ & $\alpha$\\
\hline \hline

$\sigma < 800$ km s$^{-1}$ & $18.01 \pm 0.14$ & $-1.47 \pm 0.07$ \\
$\sigma > 800$ km s$^{-1}$  & $19.27 \pm 0.09$ & $-0.66 \pm 0.09$ \\
Low $L_X$  & $17.20 \pm 0.60$ & $-1.97 \pm 0.13$ \\
High $L_X$ & $19.84 \pm 0.09$ & $-0.30 \pm 0.11$ \\
 \hline
\hline
   \end{tabular}
   \end{center}  
\end{table}

\begin{figure*}
\centering
\includegraphics[width=\textwidth]{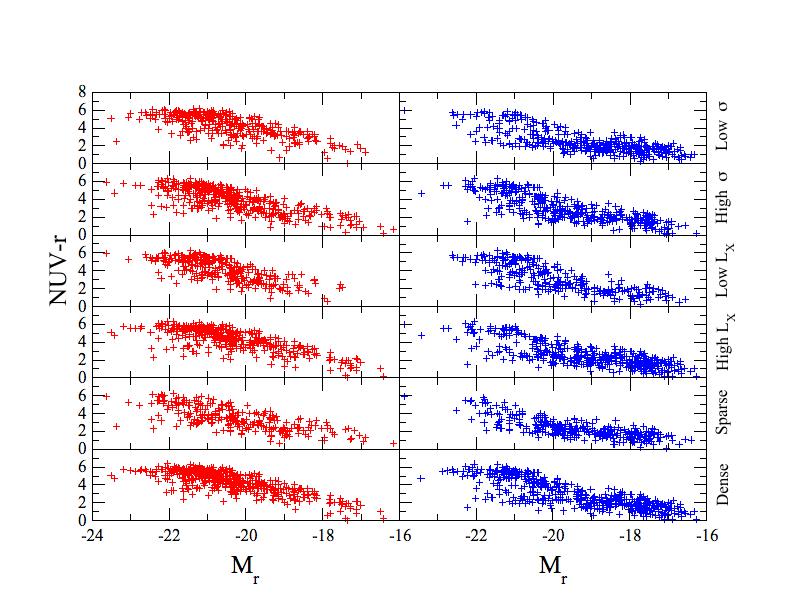}
\caption{Colour-magnitude diagrams in $NUV-r$ vs. $M_r$ for optically red (red symbols, left panels) and blue (blue symbols, right panels) galaxies in each of the cluster subsamples we consider, as indicated in the legend to the right of each subpanel. Note that the sample is selected in $NUV$ and is therefore not complete in $r$.}
\label{fig:9}
\end{figure*}

We  plot the $NUV-r$ against $M_r$ colour magnitude distributions for all the subsamples in Fig~\ref{fig:9} to understand the origin of these differences. As expected red sequence galaxies do not appear to show significant differences according to cluster environment, as most of their stellar populations are old and little significant star formation is likely to take place in these systems. Perhaps the only notable variation is that in sparse systems there appear to be a higher proportion of optically red galaxies with relatively blue $NUV-r$ and in comparison to the dense systems a smaller proportion of massive optically red galaxies with $NUV-R>5$. Together, this may indicate that some sporadic low level star formation (contributing only to the UV flux) may persist in red galaxies in sparse environments. There appear to be more significant variations for blue galaxies. The most significant is  an  excess of optically massive blue galaxies ($M_R<-20$) with red $NUV-r$ colours in comparison to the number of lower mass blue galaxies  with blue $NUV-r$ colours in clusters of low X-ray luminosity, compared to galaxies within clusters in the high X-ray luminosity subsample.  Also, in the same subsamples it appears that almost all of the blue galaxies with $M_R<-20$ and  $3<NUV-R<5$ \citep[the NUV green valley,][]{salim2014} fall in the low $L_x$ subsample. This may indicate either a slower fading of UV emission as star formation is quenched in the low $L_x$ sample or that any sporadic restarting or flickering of star formation is completely suppressed in the high $L_x$ sample, i.e. quenching in the high $L_x$ clusters allows for no restarting (even temporarily) of some star formation, presumably due to highly effective or decisive  removal/disruption of the fuelling gas supply. 

\section{Discussion}

We have derived a composite LF for 6471 NUV-selected galaxies in 28 low redshift clusters, with highly
complete redshift coverage, and examined how these composite LFs depend on cluster
properties and how red sequence and blue cloud galaxies are affected by their environment,
especially as regards star formation

We find that the cluster UV LF has $M^*$ comparable, or slightly brighter, than the field but 
has a considerably steeper faint end slope. Our results are in good agreement with measurements in 
Coma and the Shapley supercluster, but Virgo and the outer regions of Coma have flatter 
LFs and fainter $M^*$. By splitting our data into several subsamples, we argue that 
these variations are due to small number statistics when looking at the bright (and 
sparsely populated) end of single clusters.

The steeper slope in clusters appears to be due to the contribution from NUV-faint but
optically bright early-type galaxies, that are common and prominent in clusters but nearly
absent in the field (as originally argued by \citeauthor{cortese2008} \citeyear{cortese2008} 
and \citeauthor{haines2011} \citeyear{haines2011}). The LF is steeper as a result of sampling
the exponentially rising part of the galaxy mass function and because $NUV$--optical colours of quiescent 
galaxies are dominated by hot horizontal branch stars, rather than ongoing star formation,  and become bluer for higher 
luminosities (and masses) - though never as blue as for strong ongoing star formation, as shown in Fig. 5. The slope of the LF therefore may depend on the sampling depths to some 
extent (as pointed out by \citeauthor{boselli2016b} \citeyear{boselli2016b}). 

The total LF does not appear to change with velocity dispersion of the clusters but appears to have  some dependence on 
X-ray luminosity and  local density. This may favour models where interactions with the 
X-ray gas (e.g., ram pressure stripping)  affect the star formation properties of galaxies as 
they fall within clusters as both of these environmental parameters correlate with  intra cluster medium density.

For this reason we have considered red and blue galaxies separately as it may be expected that the blue population is affected more by environment than the red.. Our blue galaxy LF is similar to that determined for 
other individual clusters, showing that environmental effects are either weak or that star-forming 
galaxies within clusters are a transitional population, observed during a brief period prior to their quenching.
Therefore the apparent universality of the blue galaxy LF may be interpreted as a selection
effect as in \citet{cortese2008}. 

Unfortunately, the LFs we derive for red galaxies are quite uncertain and noisy. In all 
cases, a few bright galaxies distort the LF, so that the fit to a single Schechter function 
is quite poor. Taken at face value, the red LFs are relatively steep, but the dynamic range 
is small and the values we derive for the fit have large errors.

For blue galaxies we are able to observe that $M^*$ becomes  fainter in high $\sigma$ and high $L_X$ clusters, compared to low $\sigma$ and
low $L_X$ clusters, while the behaviouer of $\alpha$ depends on the range of $NUV_0$ that is fitted to (see figure \ref{fig:7}). In any event the ratio of UV luminous to faint galaxies is lower in the higher velocity dispersion and $L_x$ cluster subsamples (i.e. the more massive clusters).  This suggests that star formation is being suppressed in giant galaxies in these environments. Based on the NUV luminosities  and the apparent difference in fitted $M^*$ between subsamples, the star formation rates of the most vigorous star forming galaxies appears to decline by a factor 
of approximately 4 to 5   between the two groups of clusters with low and high X-ray 
luminosity. The variation we see between the high and low density subsamples may be consistent with the observation of  \cite{boselli2016b} 
that the slope of the the Virgo cluster's UV LF flattens from the periphery to the cluster core (i.e. low to high density). 

It is not clear whether velocity dispersion or X-ray luminosity is the most important 
parameter affecting ongoing star formation, but of course high-mass clusters tend to have both higher velocity dispersion 
and  X-ray luminosity, while the few high $\sigma$ clusters that are underluminous 
in X-rays may be recent mergers. However, the blue galaxies LFs for low and high X-ray 
luminosity clusters appear to differ more than the equivalent LFs for high and low 
velocity dispersion clusters, which may indicate that effects such as (gas-dependent) ram stripping may 
be more important than (galaxy-dependent) mergers and interactions. Indeed, we may expect that in clusters
with high $\sigma$ these latter effects will be less relevant, although harassment and flybys
may be more effective in these cases. 

Our data suggest that while bright red sequence galaxies are comparatively unaffected by 
the cluster environment (as their NUV emission derives from old stellar populations), 
there should be significant populations of fainter red galaxies showing the effects of 
recent quenching. Deeper FUV and NUV observations of nearby clusters with existing 
and upcoming instrumentation would be needed to confirm this hypothesis, such galaxies are potentially missing from our samples due to our NUV flux limit.

Clearly interpretation of the luminosity functions that we have presented is potentially complicated by  the range in the quality of the Schechter function fit to the various subsamples obtained by splitting the data set into two equal halves dependent of different cluster parameters. While one pair of subsamples may be well fitted,  a different split {\it of the same data} may lead to luminosity distributions that are not well fit by Schechter functions. Nevertheless, by focussing where necessary on the ratio of UV bright to faint galaxies in particular pairs of subsamples, we have been able to estrablish clear, robust differences in their luminosity functions.

\begin{acknowledgements}

The PanSTARRS1 Surveys (PS1) and the PS1 public science archive have been made possible through contributions by the Institute for Astronomy, the University of Hawaii, the PanSTARRS Project Office, the Max-Planck Society and its participating institutes, the Max Planck Institute for Astronomy, Heidelberg and the Max Planck Institute for Extraterrestrial Physics, Garching, The Johns Hopkins University, Durham University, the University of Edinburgh, the Queen's University Belfast, the Harvard-Smithsonian Center for Astrophysics, the Las Cumbres Observatory Global Telescope Network Incorporated, the National Central University of Taiwan, the Space Telescope Science Institute, the National Aeronautics and Space Administration under Grant No. NNX08AR22G issued through the Planetary Science Division of the NASA Science Mission Directorate, the National Science Foundation Grant No. AST-1238877, the University of Maryland, Eotvos Lorand University (ELTE), the Los Alamos National Laboratory, and the Gordon and Betty Moore Foundation.
This research has made use of the NASA/IPAC Extragalactic Database (NED) which is operated by the Jet Propulsion Laboratory, California Institute of Technology, under contract with the National Aeronautics and Space Administration. This research has made use of the SIMBAD database,
operated at CDS, Strasbourg, France. This research has made use of the VizieR catalogue access tool, CDS, Strasbourg, France. The original description of the VizieR service was published in A\&AS 143, 23. This research has made use of the X-Rays Clusters Database (BAX)
which is operated by the Laboratoire d'Astrophysique de Tarbes-Toulouse (LATT),
under contract with the Centre National d'Etudes Spatiales (CNES). We acknowledge the referee for his/her perceptive commentary.

\end{acknowledgements}

%
%
\bibliographystyle{aa}
\bibliography{references.bib}

\begin{appendix}

\section{Colour-magnitude relations for all clusters (Figure 1)}

\begin{figure*}
\centering
\includegraphics[width=0.9\textwidth]{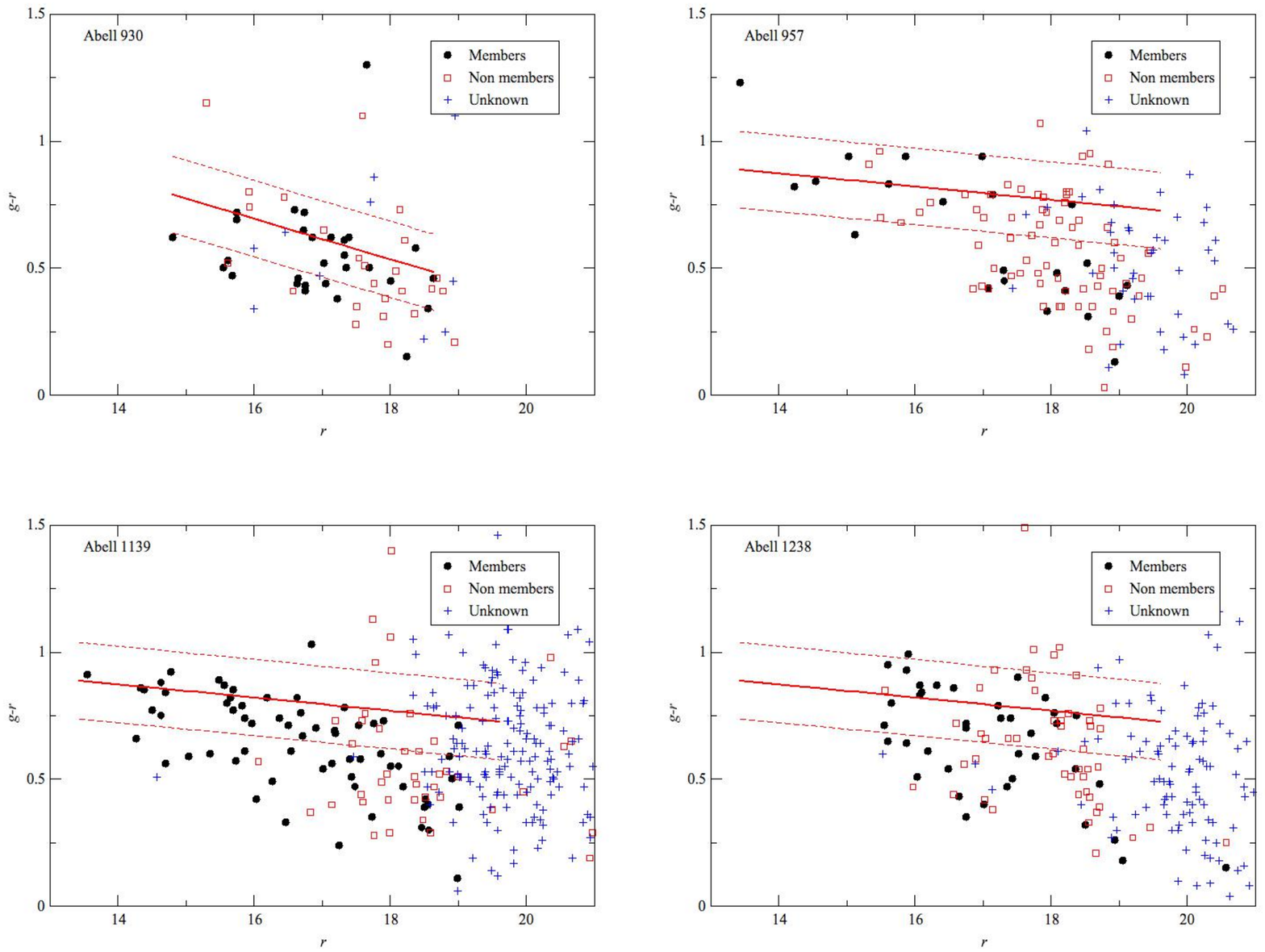}\\
\includegraphics[width=0.9\textwidth]{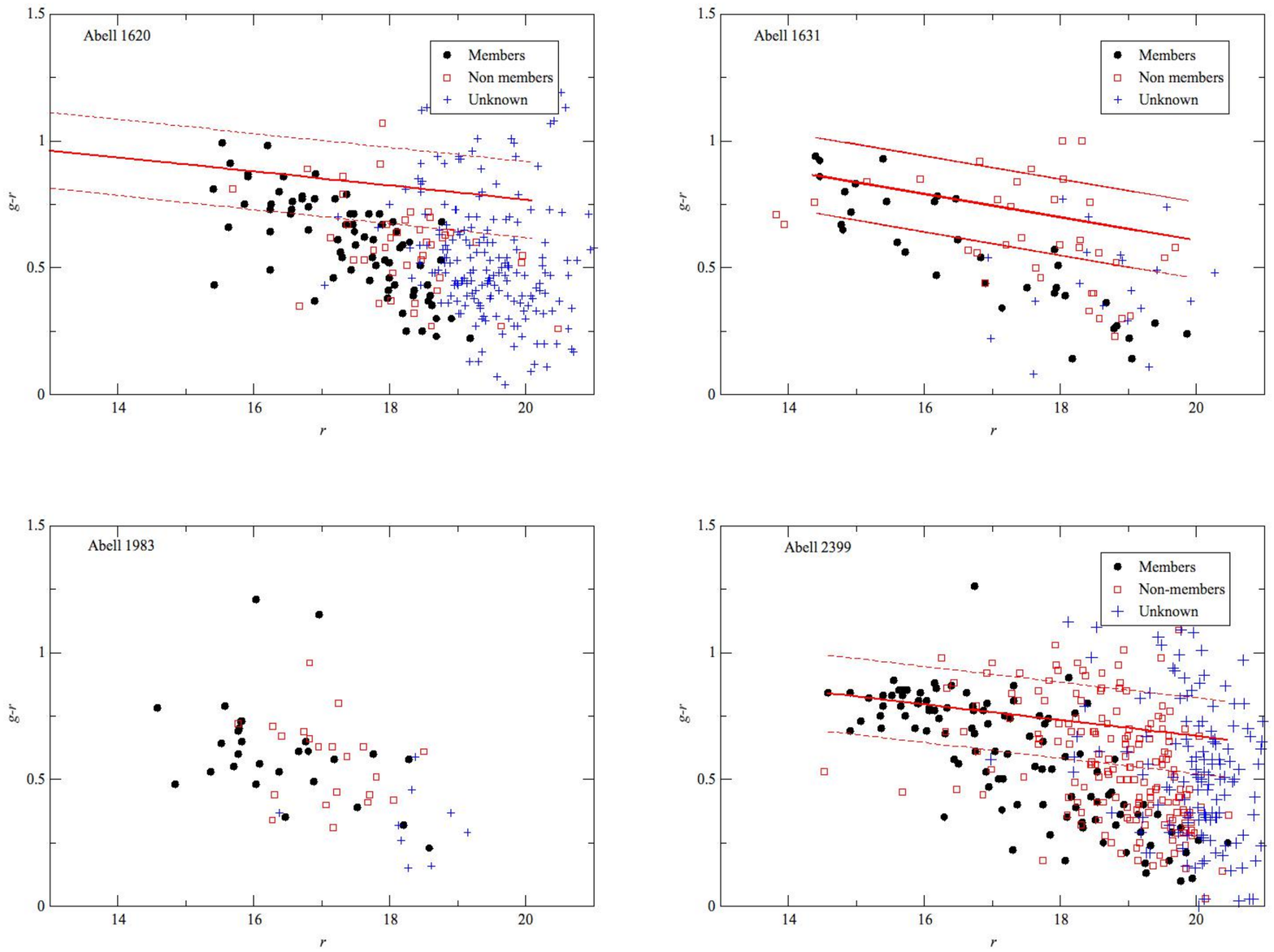}
\end{figure*}

\begin{figure*}
\centering
\includegraphics[width=0.9\textwidth]{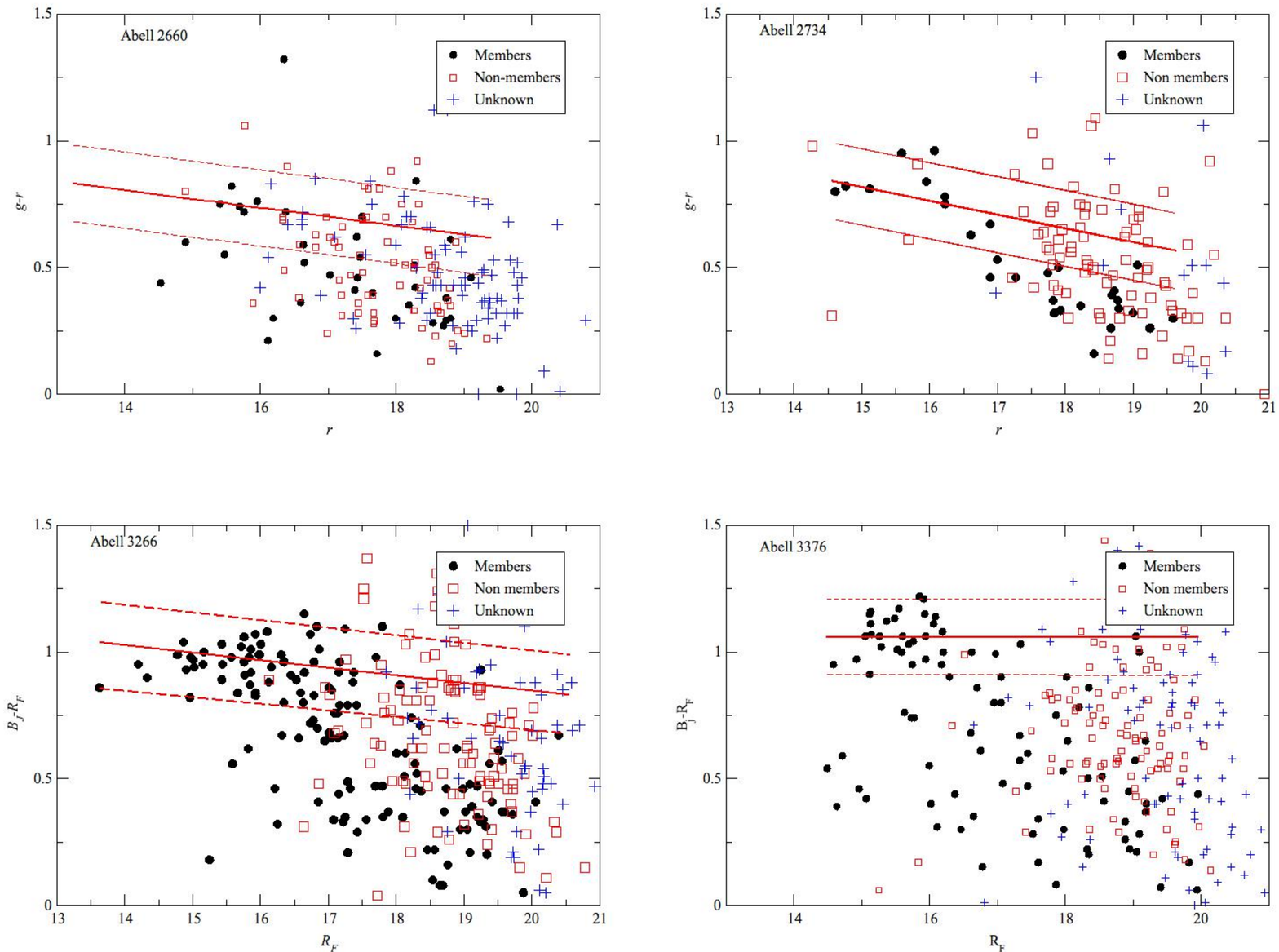}\\
\includegraphics[width=0.9\textwidth]{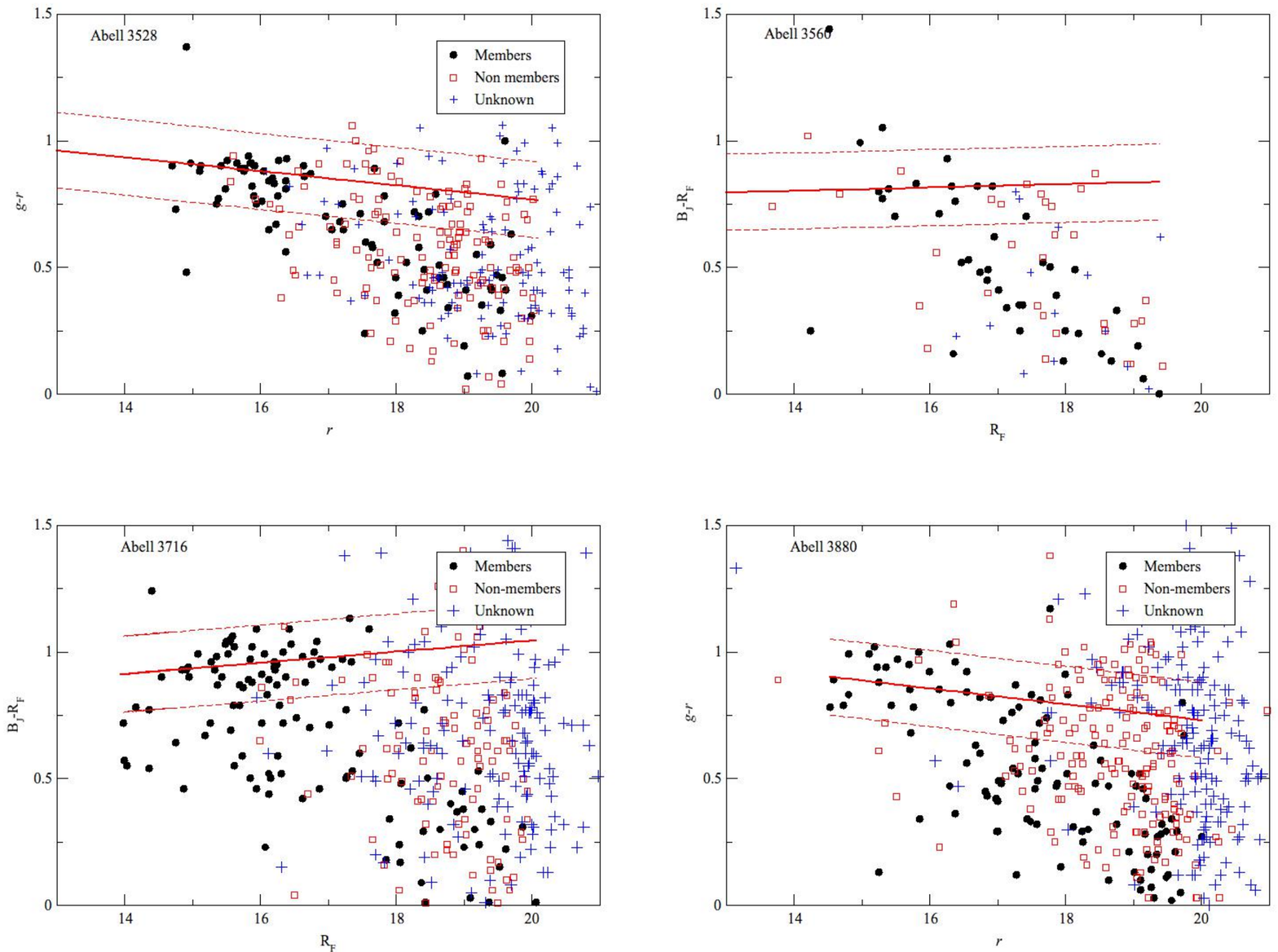}\\
\end{figure*}

\begin{figure*}
\centering
\includegraphics[width=0.9\textwidth]{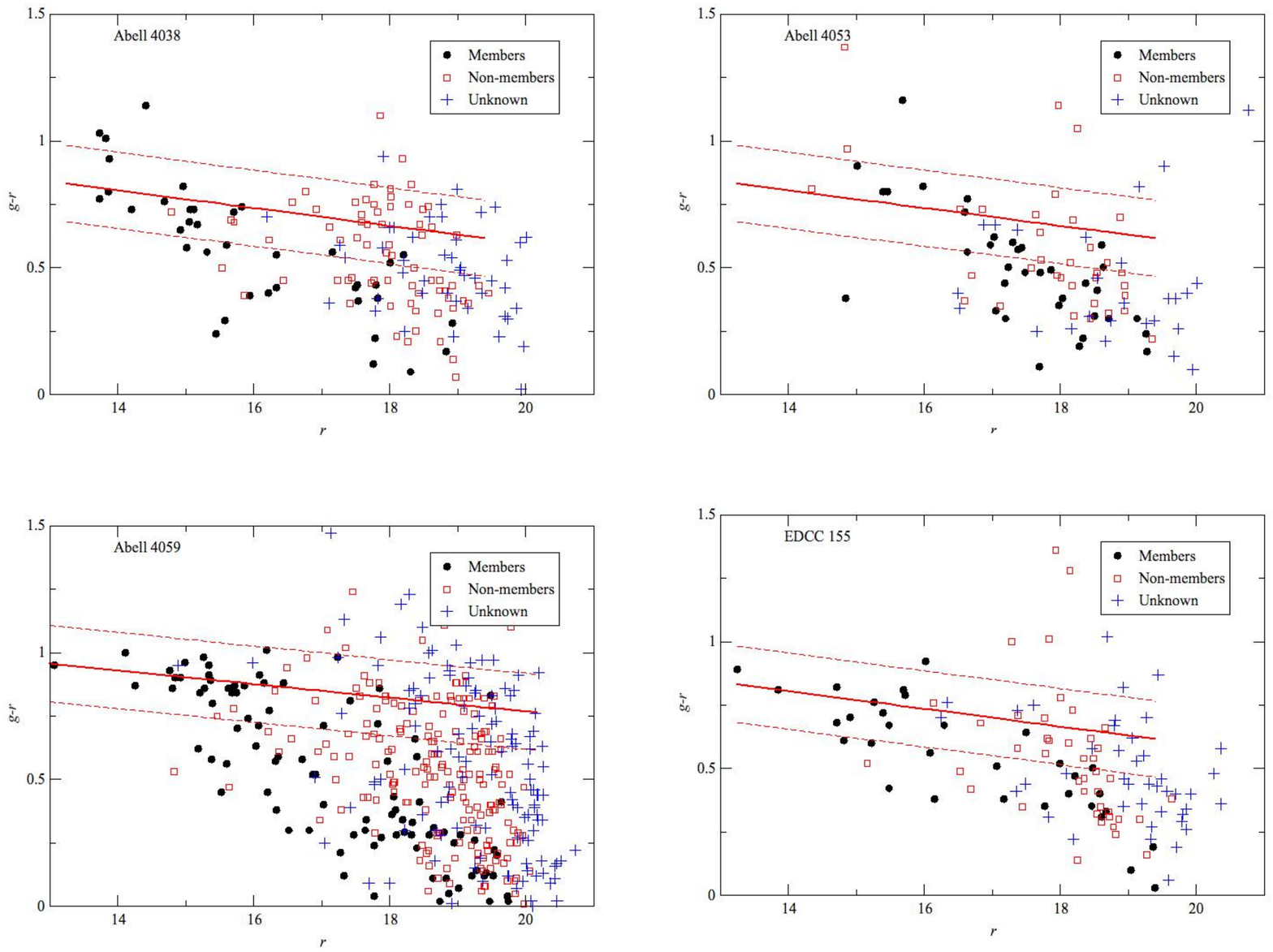}\\
\includegraphics[width=0.9\textwidth]{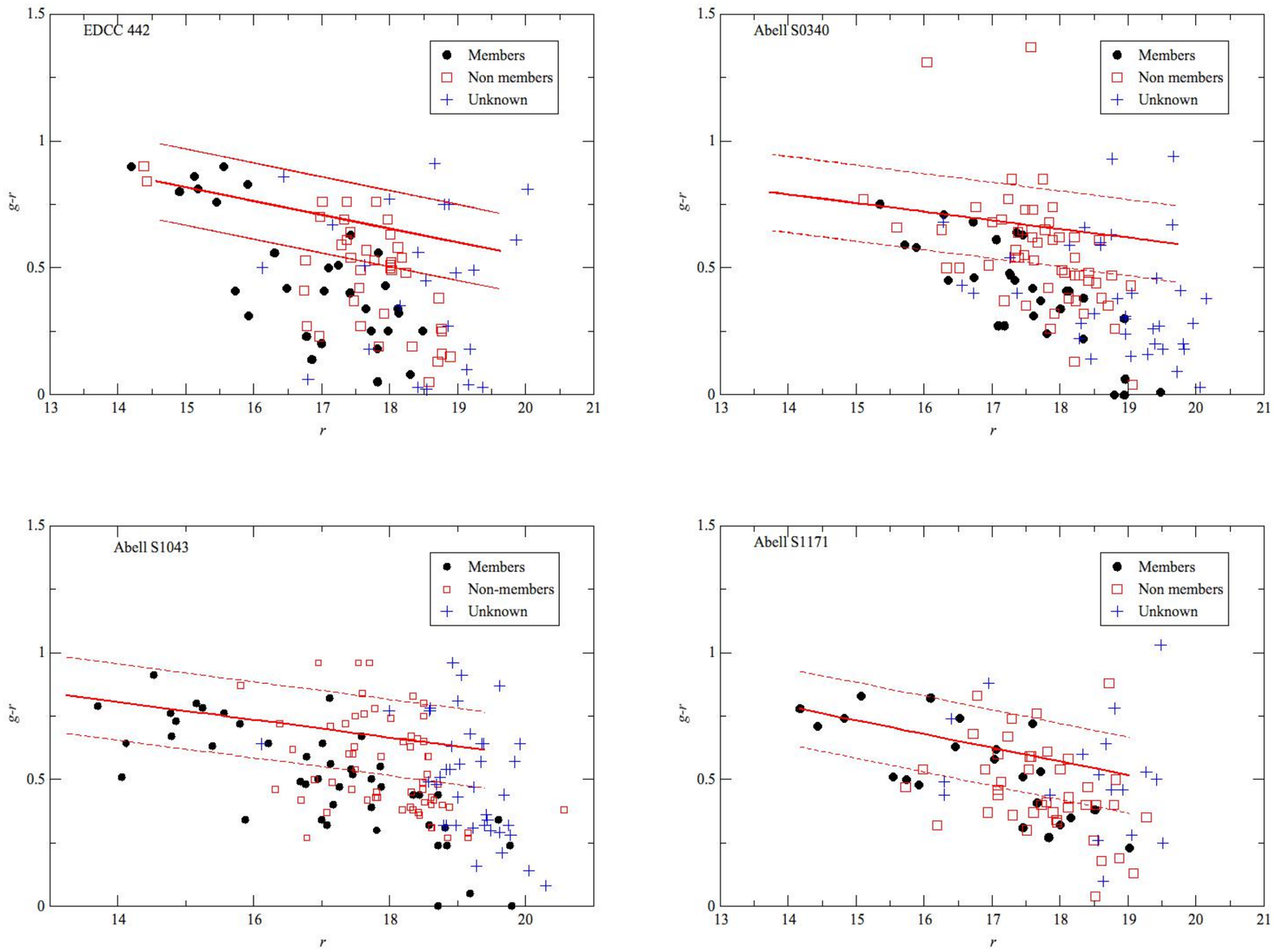}\\
\end{figure*}
\clearpage

\section{NUV colour magnitude diagrams (Figure 2)}

\clearpage

\begin{figure*}
\centering
\includegraphics[width=0.9\textwidth]{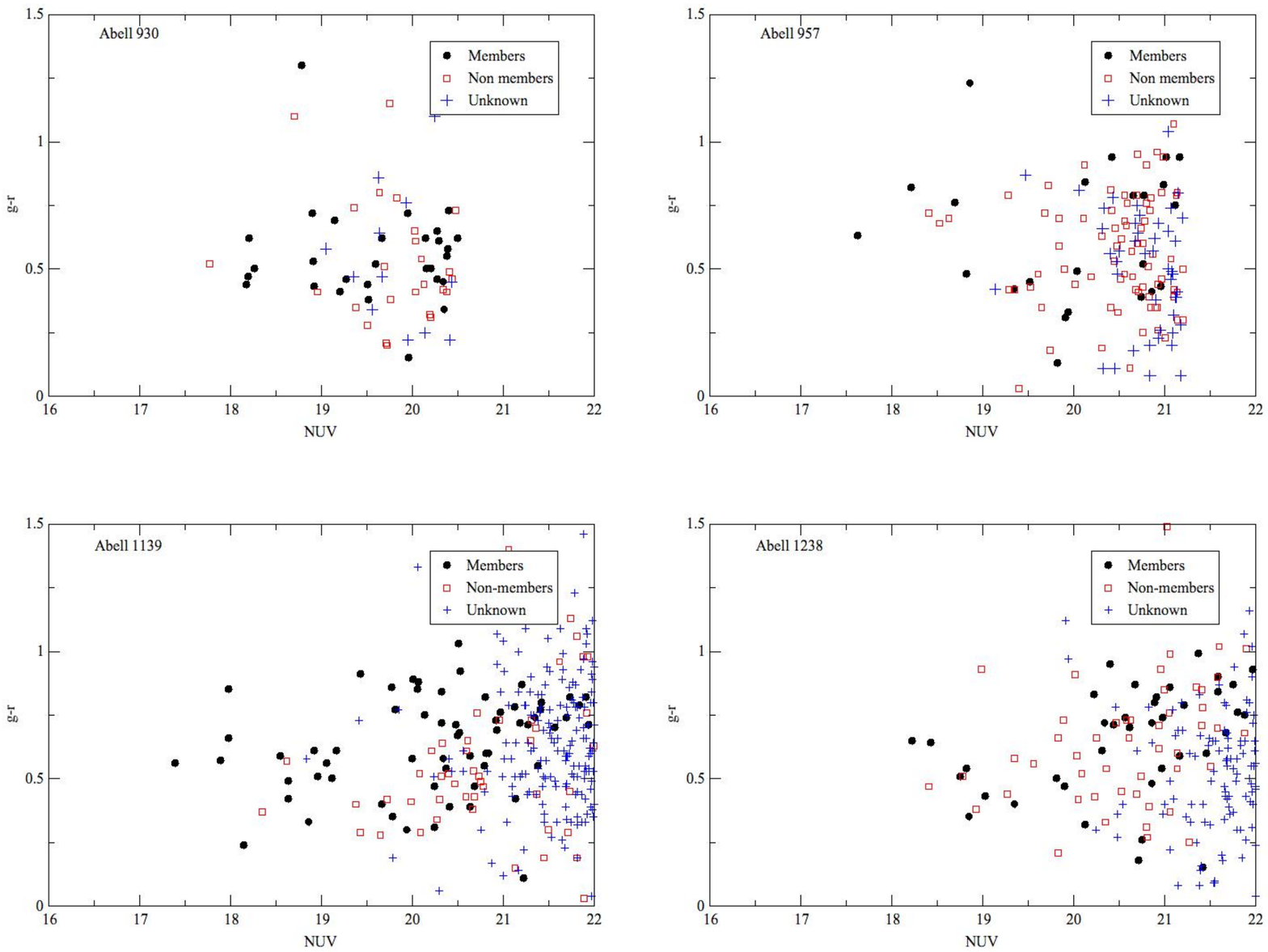}\\
\includegraphics[width=0.9\textwidth]{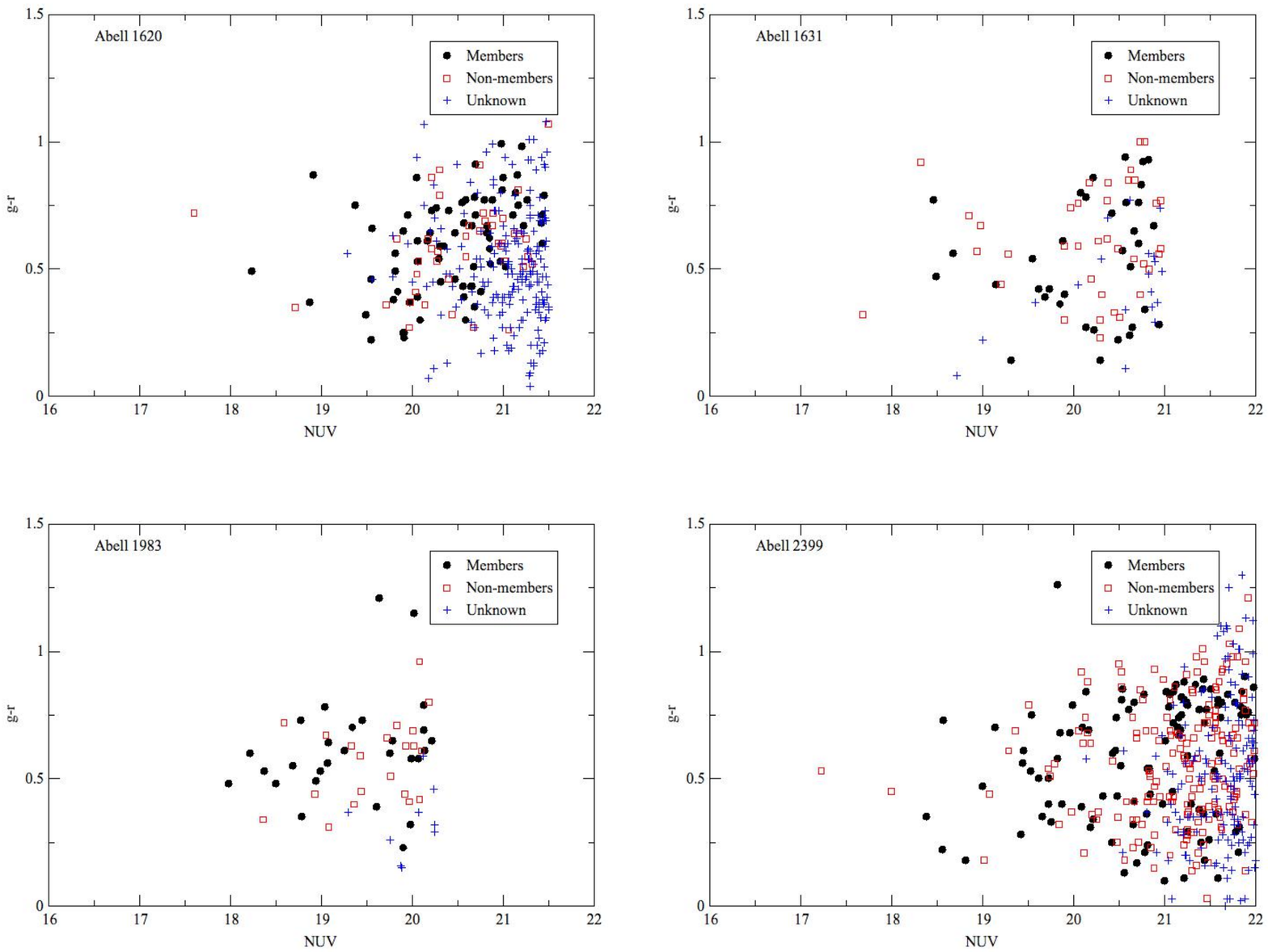}
\end{figure*}

\begin{figure*}
\centering
\includegraphics[width=0.9\textwidth]{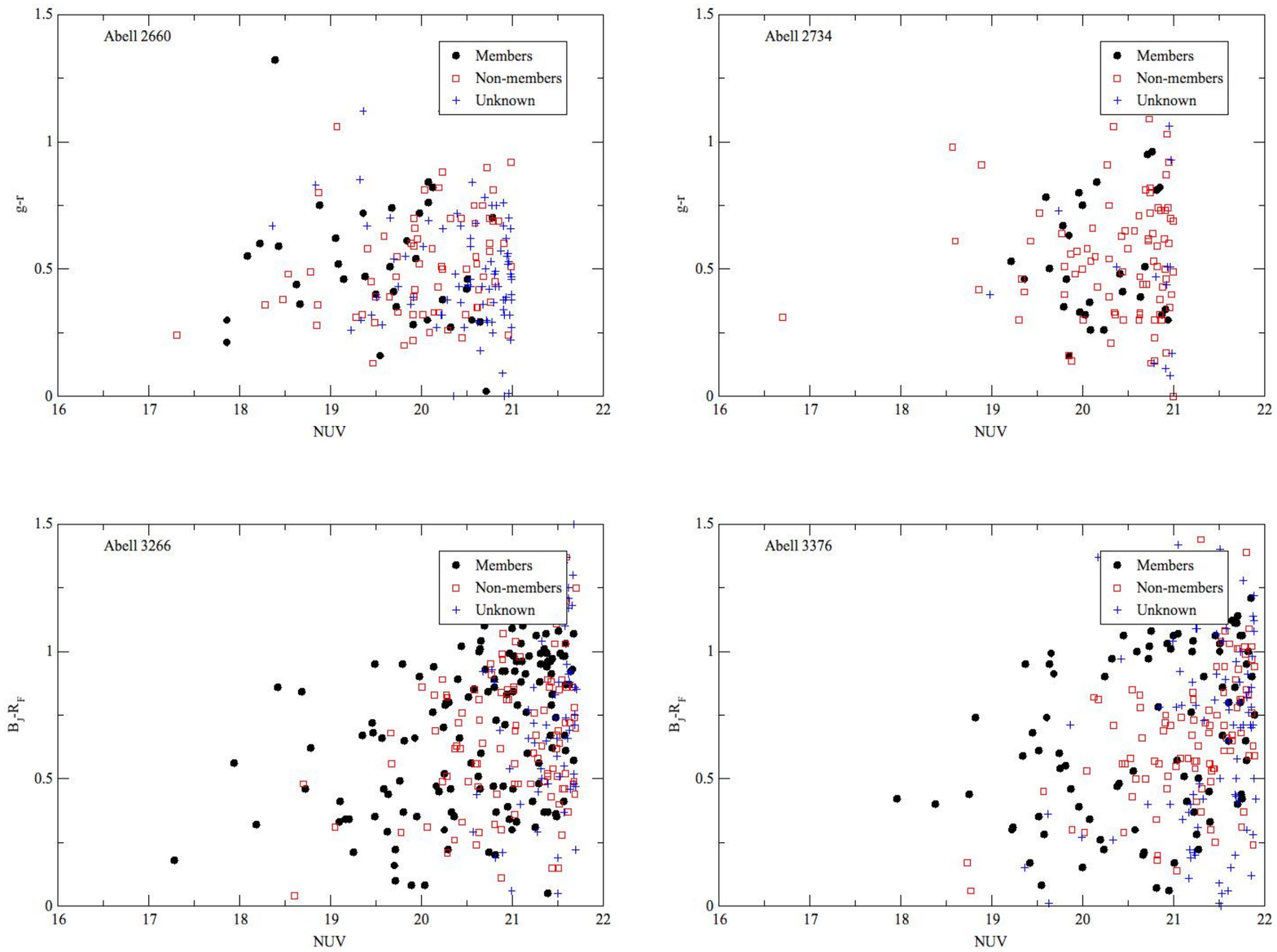}\\
\includegraphics[width=0.9\textwidth]{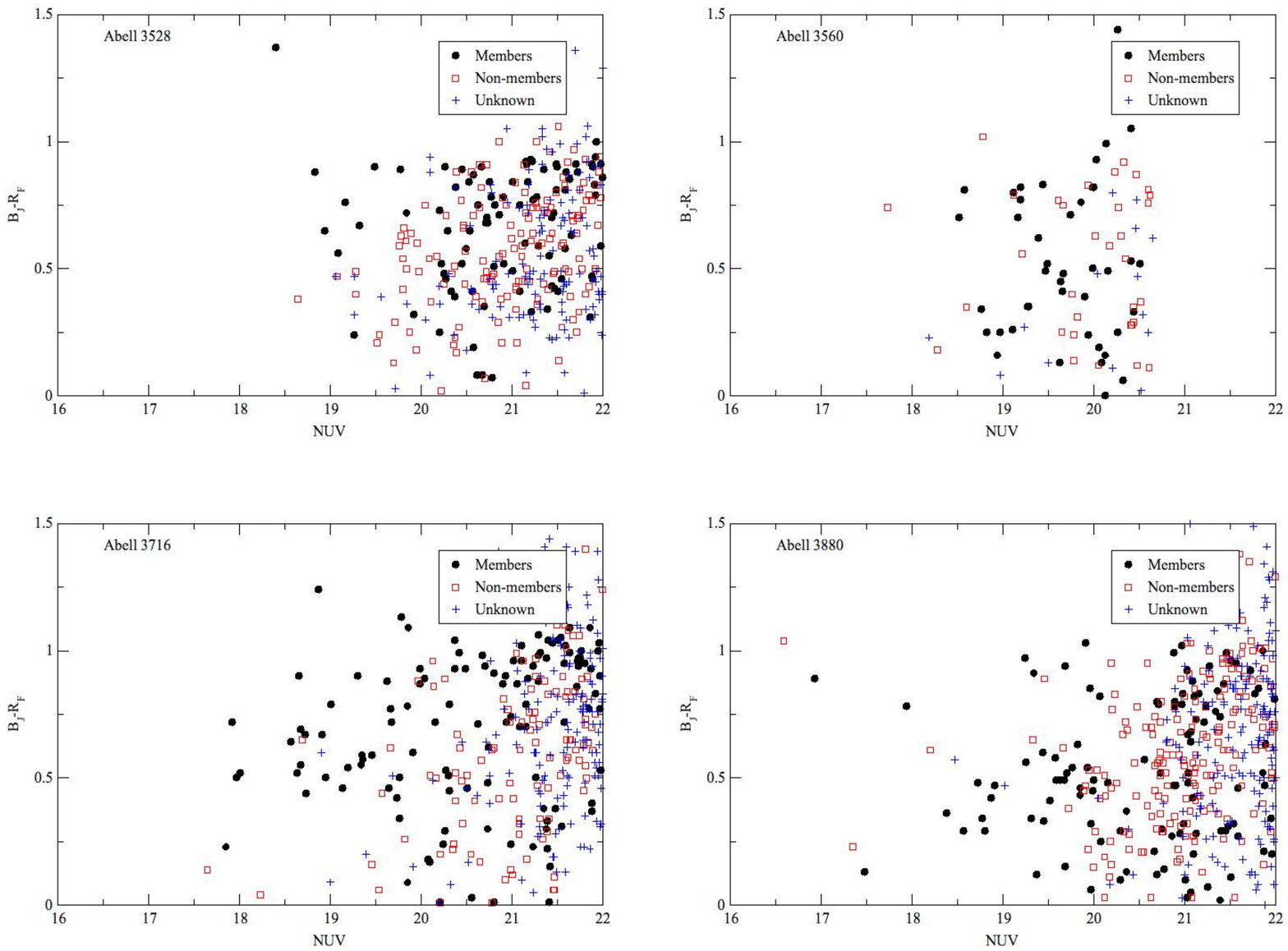}\\
\end{figure*}

\begin{figure*}
\centering
\includegraphics[width=0.9\textwidth]{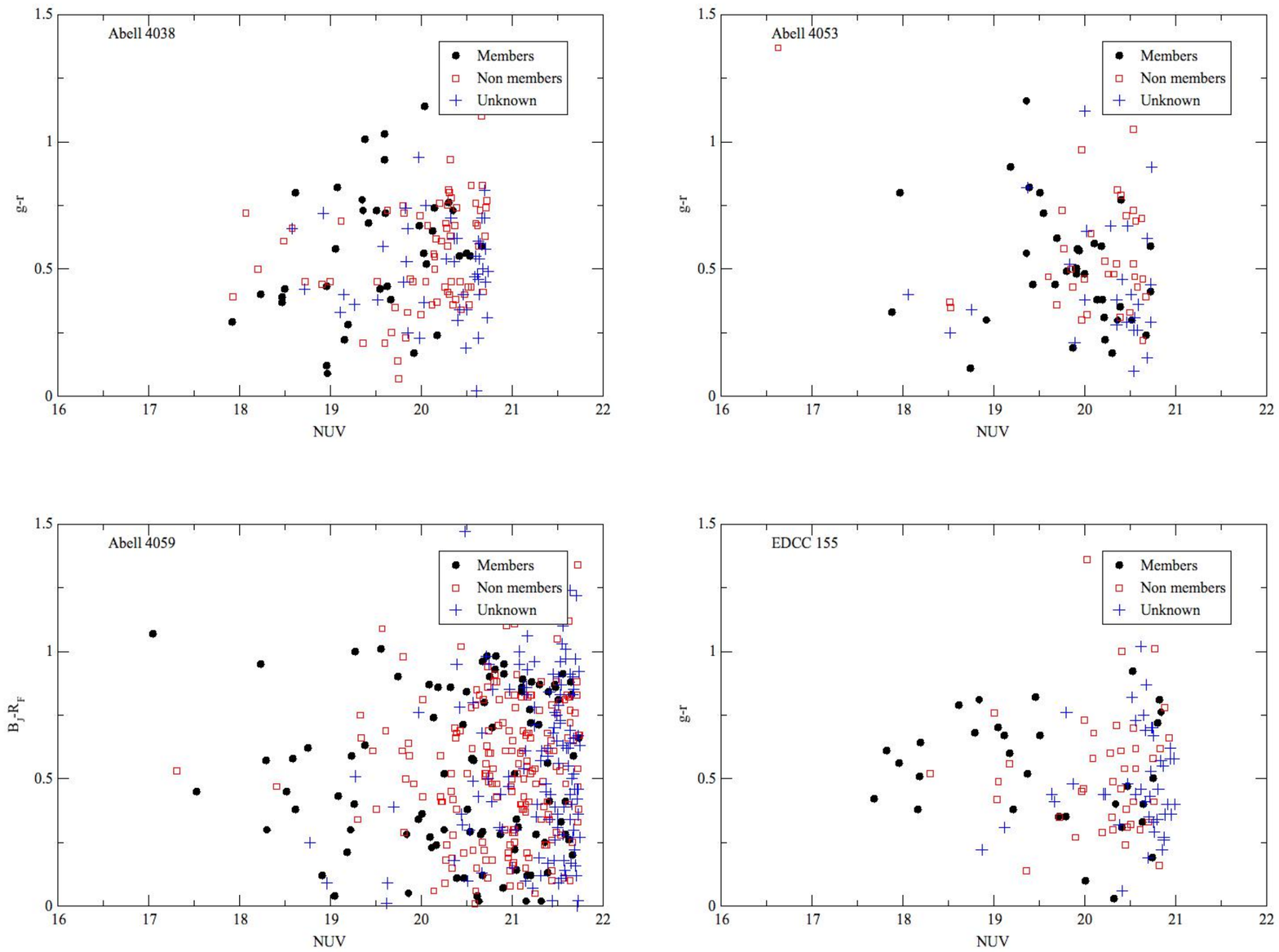}\\
\includegraphics[width=0.9\textwidth]{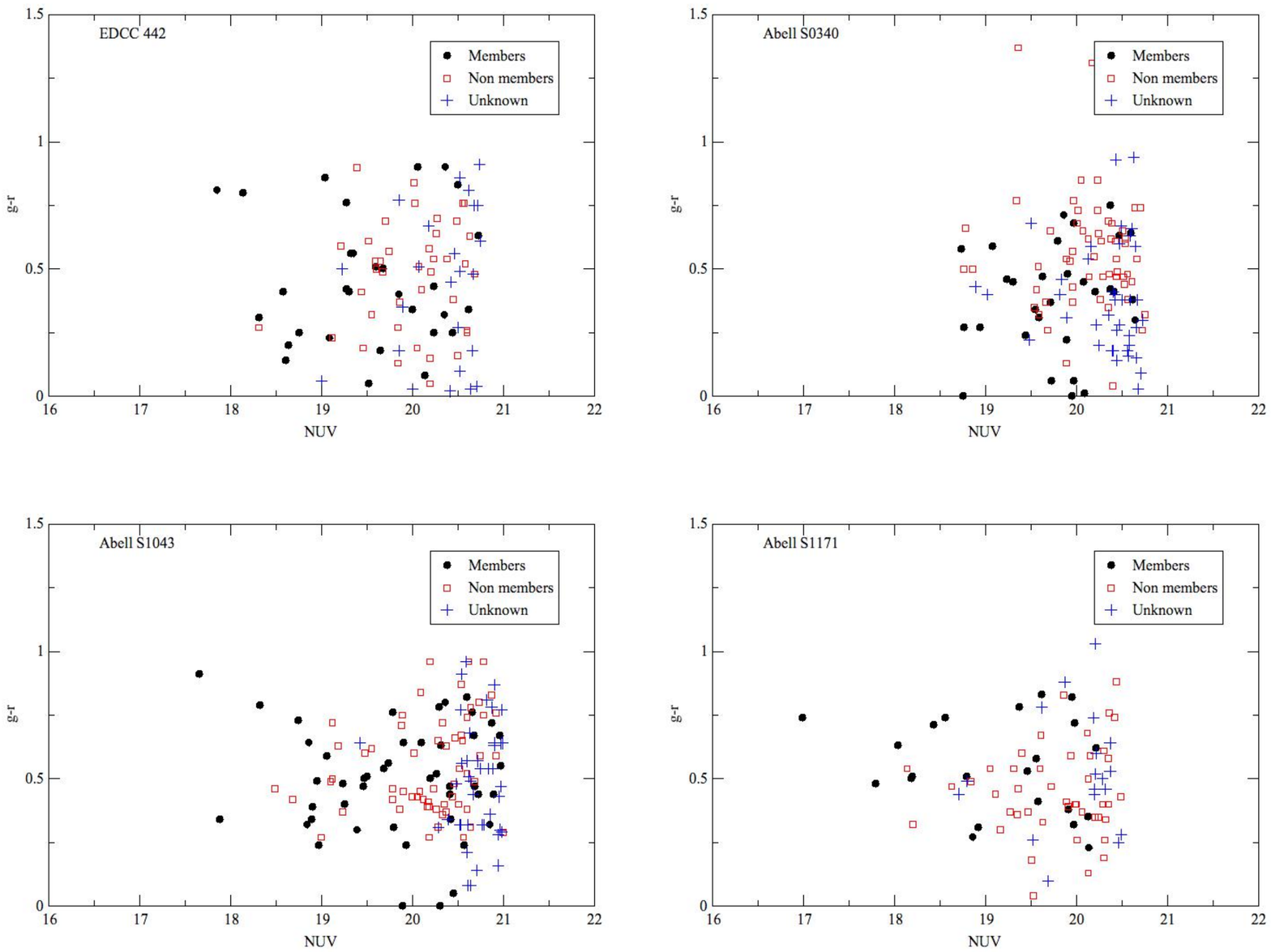}\\
\end{figure*}
\clearpage

\section{Completeness fractions for all clusters (Figure 3)}

Where the numbers refer to the order in Table 1.

\begin{figure*}
\centering
\includegraphics[width=0.9\textwidth]{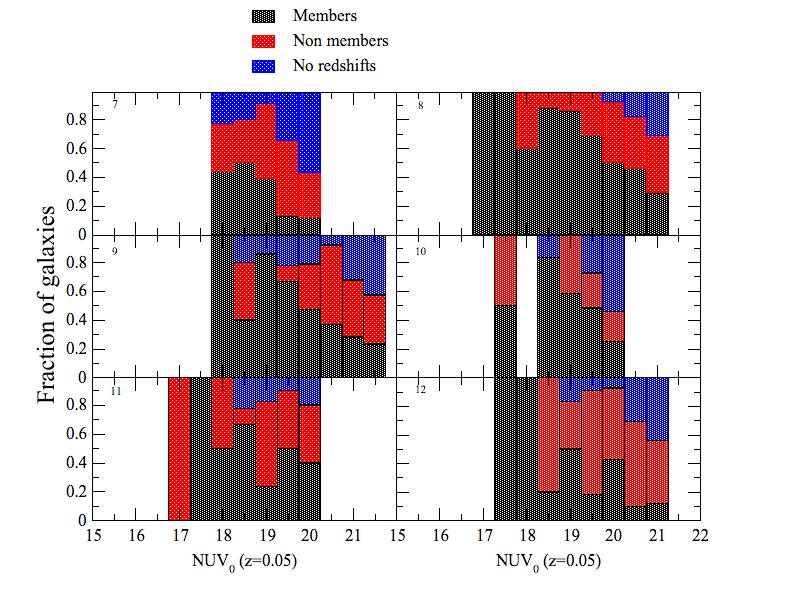}\\
\includegraphics[width=0.9\textwidth]{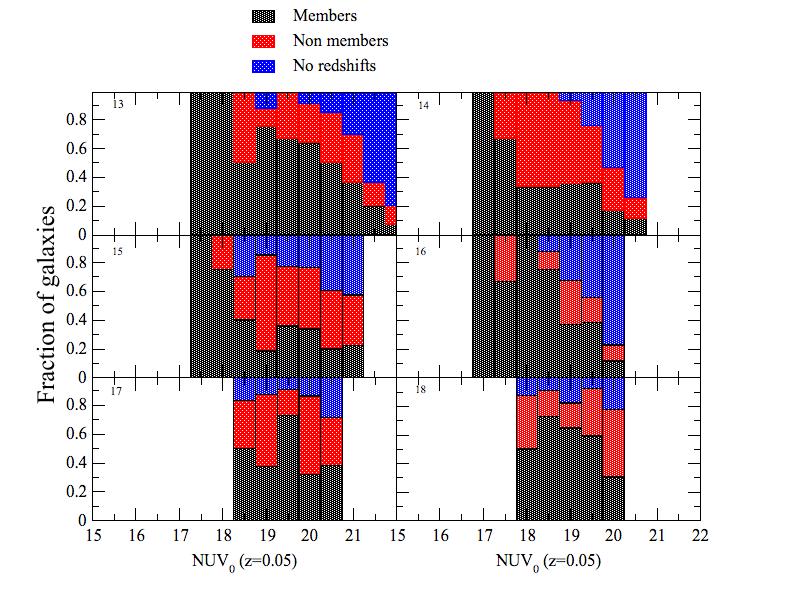}
\end{figure*}

\begin{figure*}
\centering
\includegraphics[width=0.9\textwidth]{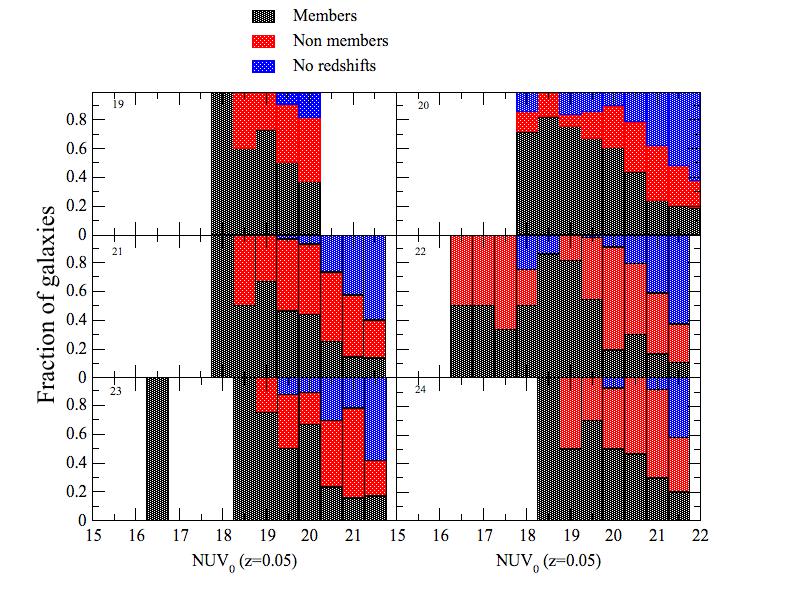}\\
\includegraphics[width=0.9\textwidth]{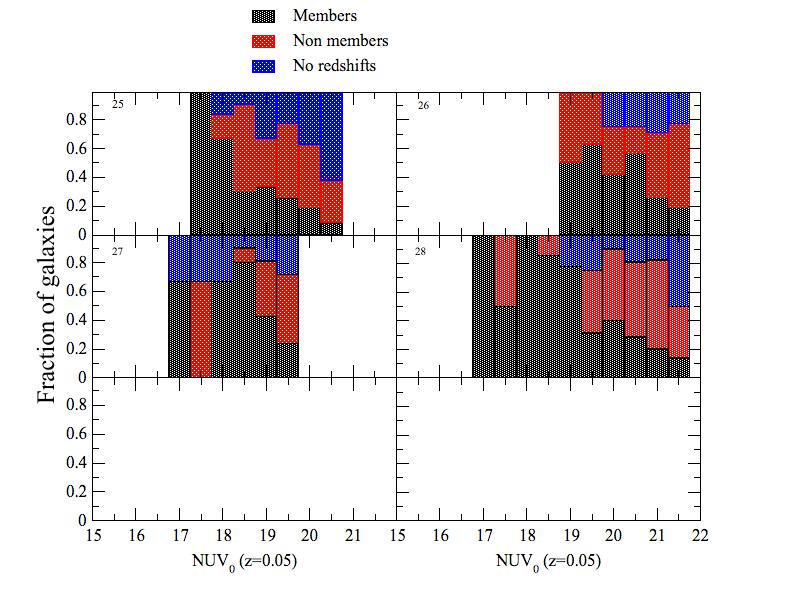}\\
\end{figure*}
\end{appendix}
\end{document}